\documentclass[11pt]{article}
\usepackage{moriond,epsfig}
\usepackage{amssymb}
\begin{document}
\vspace*{4cm}
\def\be{\begin{equation}}
\def\ee{\end{equation}}
\def\a{\alpha}
\def\b{\beta}
\def\c{\chi}
\def\d{\delta}
\def\e{\epsilon}
\def\f{\phi}
\def\g{\gamma}
\def\h{\eta}
\def\i{\iota}
\def\j{\psi}
\def\k{\kappa}
\def\l{\lambda}
\def\m{\mu}
\def\n{\nu}
\def\o{\omega}
\def\p{\pi}
\def\q{\theta}
\def\r{\rho}
\def\s{\sigma}
\def\t{\tau}
\def\u{\upsilon}
\def\x{\xi}
\def\z{\zeta}
\def\D{\Delta}
\def\F{\Phi}
\def\G{\Gamma}
\def\J{\Psi}
\def\L{\Lambda}
\def\O{\Omega}
\def\P{\Pi}
\def\Q{\Theta}
\def\S{\Sigma}
\def\U{\Upsilon}
\def\X{\Xi}


\def\ve{\varepsilon}
\def\vf{\varphi}
\def\vr{\varrho}
\def\vs{\varsigma}
\def\vq{\vartheta}

\def\ri{\rm i}
\def\rf{\rm f}

\def\dg{\dagger}                                     
\def\ddg{\ddagger}                                   
\def\wt#1{\widetilde{#1}}                    
\def\mt{\widetilde{m}_1}
\def\mb{\overline{m}}
\def\th{\tilde{h}}
\def\VEV#1{\left\langle #1\right\rangle}        
\def\beq{\begin{equation}}
\def\eeq{\end{equation}}
\def\bea{\begin{eqnarray}}
\def\eea{\end{eqnarray}}
\def\NO{\nonumber}
\def\Bar#1{\overline{#1}}

\def\pl#1#2#3{Phys.~Lett.~{\bf B {#1}} ({#2}) #3}
\def\np#1#2#3{Nucl.~Phys.~{\bf B {#1}} ({#2}) #3}
\def\prl#1#2#3{Phys.~Rev.~Lett.~{\bf #1} ({#2}) #3}
\def\pr#1#2#3{Phys.~Rev.~{\bf D {#1}} ({#2}) #3}
\def\zp#1#2#3{Z.~Phys.~{\bf C {#1}} ({#2}) #3}
\def\cqg#1#2#3{Class.~and Quantum Grav.~{\bf {#1}} ({#2}) #3}
\def\cmp#1#2#3{Commun.~Math.~Phys.~{\bf {#1}} ({#2}) #3}
\def\jmp#1#2#3{J.~Math.~Phys.~{\bf {#1}} ({#2}) #3}
\def\ap#1#2#3{Ann.~of Phys.~{\bf {#1}} ({#2}) #3}
\def\prep#1#2#3{Phys.~Rep.~{\bf {#1}C} ({#2}) #3}
\def\ptp#1#2#3{Progr.~Theor.~Phys.~{\bf {#1}} ({#2}) #3}
\def\ijmp#1#2#3{Int.~J.~Mod.~Phys.~{\bf A {#1}} ({#2}) #3}
\def\mpl#1#2#3{Mod.~Phys.~Lett.~{\bf A {#1}} ({#2}) #3}
\def\nc#1#2#3{Nuovo Cim.~{\bf {#1}} ({#2}) #3}
\def\ibid#1#2#3{{\it ibid.}~{\bf {#1}} ({#2}) #3}
\vspace{25mm}
\title{LEPTOGENESIS, NEUTRINO MIXING DATA AND THE ABSOLUTE NEUTRINO MASS SCALE
\footnote{Compendium of \cite{nove} and \cite{moriond} mostly based
on \cite{pedestrians} with some new results in 3.4, 3.8 and 3.9.}}

\author{P.~Di Bari}
\address{IFAE, Universitat Aut{\`o}noma de Barcelona, 08193 Bellaterra (Barcelona), Spain}
\maketitle

\abstracts{Recent developments in thermal leptogenesis are
reviewed. Neutrino mixing data favor a simple picture where
the matter-anti matter asymmetry is generated by the decays of the
heavy RH neutrinos {\em mildly} close to thermal equilibrium and,
remarkably, in the full non relativistic regime. This results into
predictions of the final baryon asymmetry not depending on the
initial conditions and with minimized theoretical uncertainties.
After a short outline of a geometrical derivation of the $C\!P$
asymmetry bound, we derive analytic bounds on the lightest
RH neutrino mass and on the absolute neutrino mass scale. Neutrino
masses larger than $0.1\,{\rm eV}$ are not compatible with the
minimal leptogenesis scenario. We discuss how the results get just
slightly modified within the minimal supersymmetric standard
model. In particular a conservative lower bound on the reheating temperature,
$T_R\gtrsim 10^{9}\,{\rm GeV}$,
is obtained in the relevant effective neutrino mass range $\mt\gtrsim 3\times 10^{-3}\,{\rm eV}$.
We also comment on the existence of a
`too-short-blanket problem' in connection with the possibility of
evading the neutrino mass upper bound.
}

\normalsize\baselineskip=15pt
\section{Introduction}

 Cosmic rays and CMBR observations indicate that our observable Universe is baryon asymmetric \cite{Glashow}.
Moreover the observation of the acoustic peaks in the power spectrum of CMBR \cite{WMAP},
combined with large scale structures observations \cite{SDSS}, provide a precise and robust measurement
of such an asymmetry that can be expressed in terms of the
{\em baryon to photon number ratio at the recombination time},
\be
\label{etaBCMB} \eta_{B}^{CMB}=(6.3\pm 0.3)\times 10^{-10} \, ,
\ee
in very good agreement with the latest determination from
(NACRE updated) Standard BBN and primordial Deuterium measurements
that give \cite{Cyburt} \be \eta_B^{SBBN}=(6.1\pm 0.5)\times
10^{-10} \, . \ee At the same time there is a growing evidence
that an inflationary stage  occurred during the early Universe. In
this case this would have diluted any pre-existing initial
asymmetry to a level many orders of magnitude below the measured
value, thus requiring an explanation of the observed baryon
asymmetry in terms of a dynamical generation, the aim of a model
of baryogenesis that necessitates the accomplishment of the three famous
Sakharov's conditions: $C$ and $C\!P$ violation, $B$ violation and departure
from thermal equilibrium. Within the Standard Model all three conditions are
fulfilled, yet the observed value is too large to be explained and therefore
a successful model of baryogenesis requires some new physics ingredient. A host
of models have been proposed since the first Sakharov idea
\cite{Sakharov}. Some examples of typologies of baryogenesis models are:
  Planck scale baryogenesis,
  baryogenesis from phase transitions,
  Affleck-Dine models,
  baryogenesis from black holes evaporation,
  models of spontaneous baryogenesis \cite{reviews}.

Even though {\em leptogenesis} \cite{fy} and {\em GUT baryogenesis} \cite{kt83,kt}
exhibit, from a particle physics point of view, substantial  differences,
they can be jointly regarded as two different examples belonging to the
oldest class of models of {\em baryogenesis from  heavy particle decays}.
Such a classification privileges the thermodynamical aspect
enlightening general properties that do not depend on
the specific particle physics framework.
We will thus discuss the  kinetic theory  of heavy particle decays
in the first part, while in the second part we will see how
leptogenesis is a specific remarkable example in which the
new physics ingredient is provided by the {\em seesaw mechanism}
and such that the observed baryon asymmetry is
nicely related to neutrino mixing data.

\section{Baryogenesis from heavy particle decays}

\subsection{Out-of-equilibrium decays}

Let us consider a self-conjugate heavy  ($M_X\gg M_{EW}$) particle $X$
whose decays are $C\!P$ asymmetric, in such a way that
the decaying rate into particles, $\Gamma$,  is in general different
from the decaying rate into anti-particles, $\bar{\Gamma}$,
and such that the single decay process into particles (anti-particles)
violate $B-L$ by a quantity $\Delta_{B-L}$ ($-\Delta_{B-L}$).
 The {\em $C\!P$ asymmetry parameter} is then conveniently defined as
\be
\varepsilon={\Gamma-\bar{\Gamma}\over \Gamma+\bar{\G}} \, .
\ee
 For a joint discussion of baryogenesis ($\Delta_{B-L}>0$)
and leptogenesis ($\Delta_{B-L}<0$) models, it is useful to
introduce the quantity $\widetilde{\varepsilon}=\Delta_{B-L}\,\varepsilon$.
The {\em total decay rate} $\Gamma_D=\Gamma+\bar{\Gamma}$
is the product of the total decay width,
$\Gamma_D^{\rm rest}$, times the averaged dilation factor
$\langle 1/\gamma\rangle$
\be
\Gamma_D=\Gamma_D^{\rm rest}\,\left\langle {1\over\gamma}\right\rangle \, .
\ee
Sphaleron processes, while inter-converting $B$ and $L$ separately,
leave $B-L$ unchanged \cite{sphalerons} and for this reason the kinetic equations
get much simpler if the $B-L$ evolution is tracked instead of the separate
$B$ or $L$ evolution.
Moreover it is convenient to use, as an independent variable, the quantity $z=M_X/T$ and
to introduce the {\em decay factor} $D=\Gamma_D/(H\,z)$.  Another useful choice is
to track the number of $X$ particles,
$N_X$, and the amount of the asymmetry, $N_{B-L}$, in a portion of comoving volume $R^3$
normalized in such a way to contain, averagely in ultra-relativistic
thermal equilibrium, just one $X$ particle (i.e. $N_{X}^{\rm eq}(z\ll 1)=1$).

The simplest case is when the $X$ life-time, $\tau=1/\Gamma_D^{\rm rest}$, is much longer than the
age of the Universe, $t_U=(2\,H)^{-1}$, at $z=1$, when the $X$ particles become non relativistic.
In this way decays will occur when the temperature is much below the $X$ mass and
the $X$-production from inverse decays, or other possible processes, is
Boltzmann suppressed.
 In this situation decays are the only relevant processes and
the kinetic equations for the X-abundance and the $B-L$ asymmetry
are particularly simple to be written,
\bea \label{keoed1}
{dN_X\over dz}& = & -D(z)\,N_X(z) \\ \label{keoed2}
{dN_{B-L}\over dz}& = & -\tilde{\varepsilon}\,\;{dN_X\over dz} \, ,
\eea
and solved,
\begin{eqnarray}\label{NBmL}
N_{B-L}(z) & = & N_{B-L}^{\rm i}+
\tilde{\varepsilon}\,\left[N_X^{\rm i}-N_X(z)\right] \\ \label{NBmL2}
N_X(z) & = & N_X^{\rm i}\,e^{-\int_{z_{\rm i}}^z\,dz' \,D(z')} \, .
\end{eqnarray}
The solutions can be fully described just in terms of the
{\em decay parameter}
\be\label{Kdef}
K={\Gamma_D^{\rm rest}\over H|_{z=1}} \, ,
\ee
in terms of which $D=K\,z\,\langle 1/\gamma \rangle$.
The dilation factor, averaged on the Boltzmann statistics, is simply
approximated by \cite{pedestrians}
\be
\left\langle{1\over\gamma}\right\rangle\simeq {z\over z+15/8} \, ,
\ee
a useful simple expression that makes possible to solve analytically
the integral in the Eq. (\ref{NBmL2}), yielding the result
\be\label{outN}
N_X(z)\simeq N_{X}^{\rm i}\,
e^{-K\ \left[{z^2 \over 2} - {15\,z \over 8}
+ \left({15\over 8}\right)^2\ln{\left(1+{8\over 15}z\right)}\right]} \, .
\ee
In particular the final $B-L$ asymmetry is given by
\be
N_{B-L}^{\rm f}=N_{B-L}^{\rm i}+\widetilde{\varepsilon}\,N_X^{\rm i} .
\ee
The baryon to photon ratio at recombination can then be obtained
dividing  by the number of photons at recombination $N_{\gamma}^{\rm rec}$
(about thirty times the number of photons at the onset of $X$ decays)
and taking into account that sphalerons will convert only a fraction
$a_{\rm sph}\simeq 1/3$ of the $B-L$ asymmetry into a baryon asymmetry.
In this way one can write:
\be\label{etaB}
\eta_B \simeq {1\over 3}\,{N_{B-L}^{\rm f}\over N_{\gamma}^{\rm rec}} \, .
\ee
\begin{figure}[t]
\centerline{\psfig{file=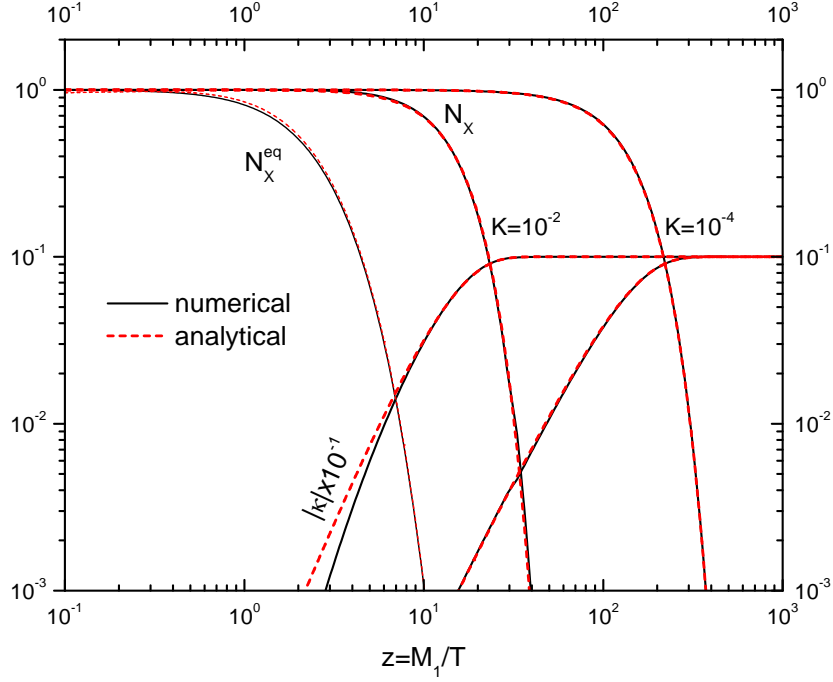,height=11cm,width=14cm}}
\caption{out of equilibrium decays. }
\end{figure}
It is useful to introduce the {\em efficiency factor} defined as
the ratio of the asymmetry produced from the $X$ decays, excluding the
contribution from a possible initial quantity, to the $C\!P$ asymmetry, i.e.
\be
\kappa(z)\equiv {N_{B-L}(z)|_{N_{B-L}^i=0}\over
\tilde{\ve}} \, .
\ee
In the case of out of equilibrium decays one
has $\k(z)=N^{\rm i}_X-N_X(z)$ and the Eq. (\ref{NBmL}) can be re-casted as
\be
N_{B-L}(z)=N^i_{B-L}+\widetilde{\ve}\,\kappa(z) \, .
\ee
This definition is such that the {\em final efficiency factor},
$\kappa_{\rm f}\equiv \k(\infty)=N_X^i$, is equal to unity in the
case of an initial thermal abundance with $z_i\ll 1$. In Fig. 1 we
show two examples of out of equilibrium decays, for $K=10^{-2}$ and
$K=10^{-4}$, assuming an initial thermal $X$ abundance ($N_X^{\rm i}=1$)
and zero initial asymmetry ($N_{B-L}^i=0$). The numerical results
are compared with the analytic expression (cf. (\ref{outN})).

 The out-of-equilibrium picture is an efficient way to produce an asymmetry from decays.
However it relies on the possibility that an initial $X$ abundance
was thermalized by some unspecified mechanism at $T\gtrsim M_X$ and that
one can neglect a possible $N_{B-L}^i$ generated during or after
inflation and before the onset of $X$ decays. Therefore,
it is evident that this picture is plagued by a
{\em strong sensitivity to the initial conditions} and hence it requires
to be complemented with a  model able to specify them, for example
a detailed description of the inflationary stage.

\subsection{Inverse decays}

The out-of-equilibrium picture is strictly valid only in the limit $K\rightarrow 0$.
If one defines $z_d$ as the value $M_X/T_D$ such that the $X$ life time
coincides with the age of the Universe ($\tau=t_U(z_D)$) then, for $K\ll 1$,
one has $z_d \simeq \sqrt{2/K}$. Thus for $K\gtrsim 1$ the $X$'s will decay
for $T_d=M_X/z_d \gtrsim M_X/\sqrt{2}$ and the inverse decays have to be
taken into account. The kinetic equations
(\ref{keoed1})  and (\ref{keoed2})
are then generalized in the following way \cite{kt,luty,plum,bcst,bdp1}
\footnote{The equations (\ref{ek1}) and (\ref{ek2}) are actually not only
accounting for decays and inverse decays but also for the
real intermediate state contribution from $2\leftrightarrow 2$
scattering processes. This term exactly cancels a $C\!P$ non conserving
term from inverse decays that would otherwise lead to an
un-physical asymmetry generation in thermal equilibrium  \cite{dolgov}.}
\bea\label{ek1}
{dN_X\over dz}& = & -D(z)\,N_X(z)+D(z)\,N_X^{\rm eq}(z) \\ \label{ek2}
{dN_{B-L}\over dz}& = & -\tilde{\varepsilon}\,\;{dN_X\over dz}-
W_{ID}(z)\,N_{B-L}(z) \, .
\eea
In the equation for $N_X$ the second term accounts for the inverse decays
that, remarkably, can now produce the $X$'s. On the other hand one can see that
a new term appears in the second equation for the asymmetry too, a
wash-out term that tends to destroy what is generated from the decays.
This term is controlled by the  (inverse decays) wash-out factor given by
\be
W_{ID}={m\over 2}\,D\,{N_X^{\rm eq}\over N_{b,l}^{\rm eq}} \propto K \, ,
\ee
where $m$ is the number of baryons or leptons in the $X$ decay final state
($m=1$ in the case of leptogenesis).
Note that the decay parameter $K$ is still the only parameter in
the equations and thus the solutions will still depend only on $K$.
They can be again worked out in an integral form \cite{kt}. In the
case of the $B-L$ asymmetry one can write the final asymmetry as
\be\label{NBmLf}
N_{B-L}^{\rm f}=N_{B-L}^{\rm i}\,e^{-\int_{z_{\rm i}}^{\infty}\,dz'\,W_{ID}(z')}
+\tilde{\varepsilon}\,\k_{\rm f} \, ,
\ee
where now the efficiency factor is given by the integral
\be\label{kf}
\kappa_{\rm f}(K,z_{\rm i})= -\int_{z_{\rm i}}^{\infty}
\,dz'\,\left[{dN_{X}\over dz'}\right]\,
\,e^{-\int_{z'}^{\infty}\,dz''\,W_{ID}(z'')} \, .
\ee
In the limit $K\rightarrow 0$ the out-of-equilibrium case is recovered.
In general one can see that the wash-out has the positive effect to
damp a pre-existing asymmetry but also the negative one to damp the same
asymmetry generated from decays, thus reducing the efficiency of the mechanism.
A quantitative analysis is crucial and it is very useful to discuss separately the
regime of strong wash out for $K\gtrsim 1$ and the regime  of weak wash-out for $K\lesssim 1$.

\subsection{Strong wash-out regime}

The strong wash-out regime is characterized by the existence,
for $K\gtrsim 3$, of an interval $[z_{\rm in},z_{\rm out}]$
such that $W_{ID}\gtrsim 1$ and thus such that inverse decays are in equilibrium.
Practically all the asymmetry produced at $z>z_{\rm out}$
is washed-out including, remarkably, an initial one. Moreover
the calculation of the residual asymmetry is made very simple by the
possibility to use the {\em close equilibrium approximation} given by
\be\label{approx}
{dN_{X}\over dz'}\simeq {dN_{X}^{\rm eq}\over dz}=-{2\over K\,z}\,W_{ID}(z) .
\ee
In this way the integral in the Eq. (\ref{kf}) can be easily evaluated \cite{kt,pedestrians}.
Indeed, this can be regarded as a Laplace integral, that means an integral of the form
\be\label{psi}
\int_0^{\infty}\,dz'\,e^{-\psi(z',z)} \, ,
\ee
that receives a dominant contribution only from a small interval centered
around a special value $z_B$ such that $d\psi/dz=0$.
In this way one can use the approximation of replacing
$W_{ID}(z'')$ with  $W_{ID}(z'')\,z_B/z''$ in  the Eq. (\ref{kf}).
With this approximation and assuming $z_i\ll 1$
the integral can be easily solved obtaining
\be\label{kfth}
k_f \simeq {2\over m\,K\,z_B}\,\left(1-e^{-{m\,K\,z_B\over 2}}\right) \, .
\ee
\begin{figure}[t]
\centerline{\psfig{file=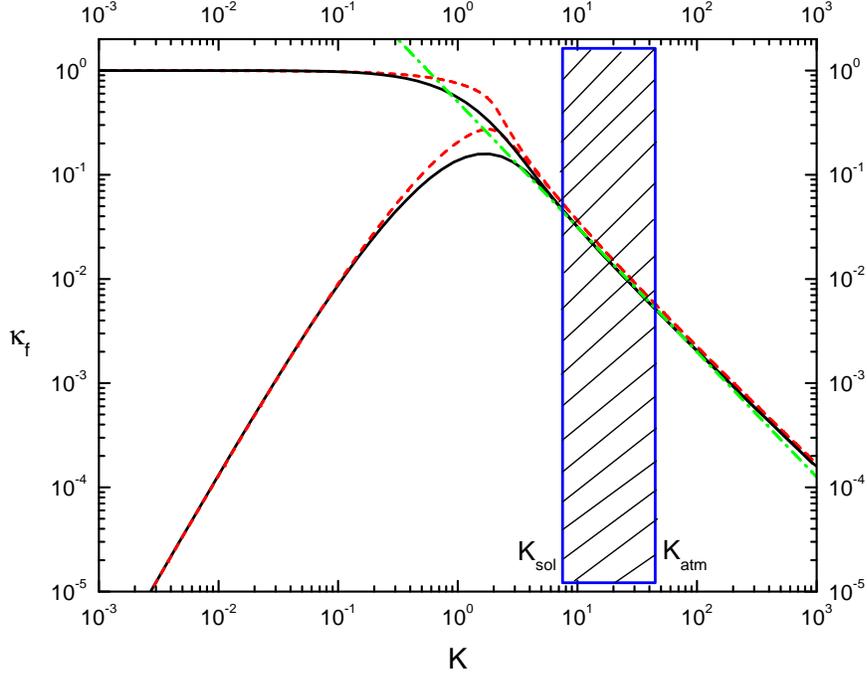,height=11cm,width=14cm}}
\caption{final efficiency factor  as a function of the decay parameter
$K$ for thermal and dynamical initial $X$ abundance. The solid lines
are the numerical solutions of the Eq.'s (\ref{ek1}) and (\ref{ek2}),
the short-dashed lines are the analytic results
(cf. (\ref{kfth}) and (\ref{kf-})+(\ref{kf+})), the dot-dashed line
is the power law fit Eq. (\ref{kfpllep}). The dashed box
is the range of values for $K$ favored by neutrino mixing data
in the case of leptogenesis (cf. (\ref{Kleprange})).}
\end{figure}
For large $K\gg 1$ and for $m=2$
this expression coincides with that one found in \cite{kt}
\footnote{Note however that the definition (\ref{Kdef}) for $K$
has to be used instead of $K=(1/2)\,(\Gamma_D/H)_{z=1}$.}.
The calculation of the important quantity $z_B$
proceeds from its definition, $(d\psi/dz)_{z_B}=0$,
approximately equivalent to the equation
\be\label{zB}
W_{I\!D}(z_{\rm B}) = \left\langle {1\over\gamma}\right\rangle^{-1}(z_{\rm B})\,
-\, {3\over z_{\rm B}}\;.
\ee
This is a transcendental algebraic equation and thus one cannot find an exact
analytic solution (see \cite{pedestrians} for an approximate
procedure). However the expression
\be\label{fit}
z_B(K)\simeq 1+4.5\,(m\,K)^{0.13}\,e^{-{5\over 2\,m\,K}} \, ,
\ee
provides quite a good fit that can be plugged into the Eq. (\ref{kfth}) thus
getting an analytic expression for the efficiency factor.
At vary large $K$ this behaves as a power law $\kappa_{\rm f}\propto K^{-1.13}$.
In Fig. 2 we compare the analytic solution for $\k_{\rm f}$ (cf. (\ref{kfth}))
with the numerical solution (for $m=1$). One can see how for $K\gtrsim 4$
the agreement is quite good.
Note that the Eq. (\ref{zB}) implies that  for large values
of K one has $z_B\simeq z_{\rm out}$, that particular value of
$z$ corresponding to the last moment when inverse decays are in equilibrium
($W_{ID}\geq 1$).  In this way almost all the asymmetry produced for
$z\lesssim z_B$ is washed-out and most of the surviving asymmetry is produced
in the period just around the inverse decays freeze out, simply because the
$X$ abundance gets rapidly Boltzmann suppressed. An example of
this picture is illustrated in Fig. 3 for $K=100$ (from \cite{pedestrians}).
\begin{figure}[t]
\centerline{\psfig{file=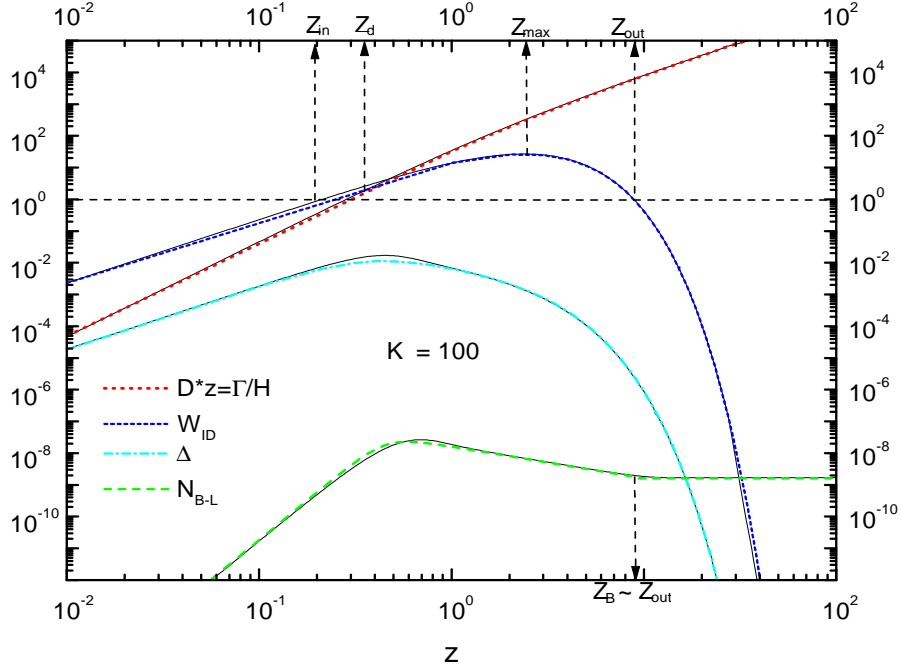,height=10cm,width=15cm}}
\vspace{-5mm}
\caption{\small comparison between analytical (short-dashed lines)
and numerical (solid lines) results in the case of
strong wash-out ($K=100$) for $|\ve_1| = 0.75\times 10^{-6}$.}
\label{strong}
\end{figure}
Instead of the abundance
we plotted the deviation from the equilibrium value, the quantity
$\Delta=N_{X}-N_{X}^{\rm eq}$. The deviation grows until the $X$'s decay
at $z\simeq z_d$, when it reaches a maximum, and decrease afterwards when
the abundance stays close to thermal equilibrium. Correspondingly the asymmetry
grows for $z\lesssim z_d$, reaching a maximum around $z\simeq 1$, and then
it is washed-out until it freezes at $z_B\simeq z_{\rm out}$.
The evolution of the asymmetry $N_{B-L}(z)$ can induce the
wrong impression that the residual asymmetry is some fraction
of what was generated at $z\simeq 1$ and that one cannot relax the
assumption $z_i \ll 1$ without reducing considerably the final value of the asymmetry.
Actually what is produced is also very quickly destroyed.
A plot of the quantity $\psi(z,\infty)$, as defined in the Eq. (\ref{psi})
and shown in Fig 4 (from \cite{pedestrians}), enlightens some interesting aspects.
\begin{figure}[t]
\centerline{\psfig{file=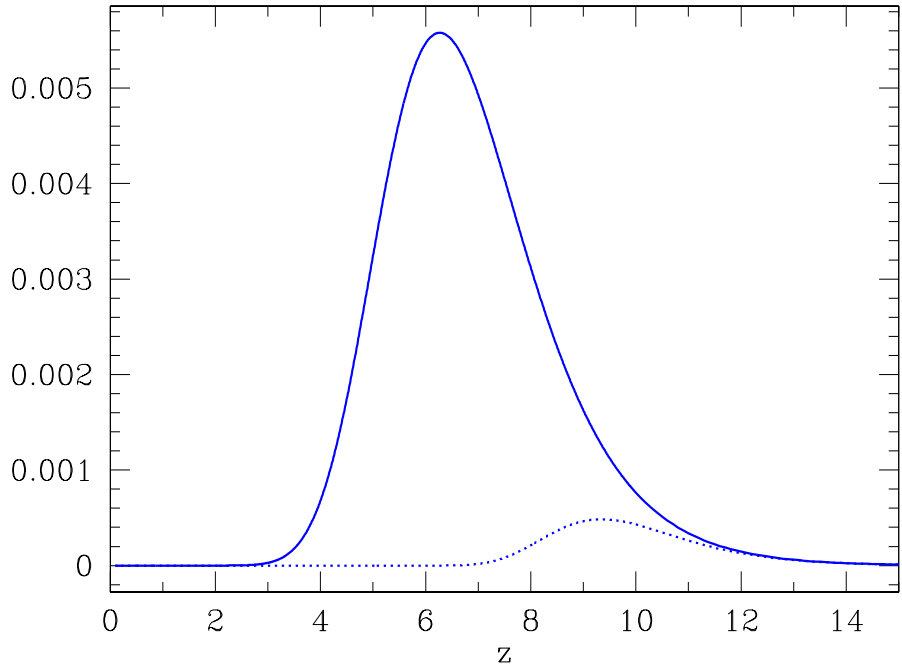,height=9cm,width=12cm}}
\vspace{-5mm}
\caption{\small the function $\psi(z',\infty)$
for $K=10$ (solid line) and $K=100$ (dashed line).}
\label{strong}
\end{figure}
This is the final asymmetry that was produced in a infinitesimal
interval around $z$. It is evident how just the asymmetry that was produced
around $z_B$ survives and, for this reason, the temperature $T_B=M_X/z_B$
can be rightly identified as the {\em temperature of baryogenesis}
for these models.
It also means that in the strong wash out regime the final
asymmetry was produced when the $X$ particles were
fully non relativistic implying that the simple kinetic
equations  (\ref{ek1}) and (\ref{ek2}), employing
the Boltzmann approximation, give actually accurate results
and corrections from use of the exact quantum statistics can be safely
neglected.

This is not the only nice feature of the strong wash-out regime.
Since any asymmetry generated for $z\lesssim z_B$ gets washed-out,
one can also rightly neglect any pre-existing initial asymmetry $N_{B-L}^i$.
At the same time the final asymmetry does not depend on the initial
$X$ abundance.
\begin{figure}[t]
\centerline{\psfig{file=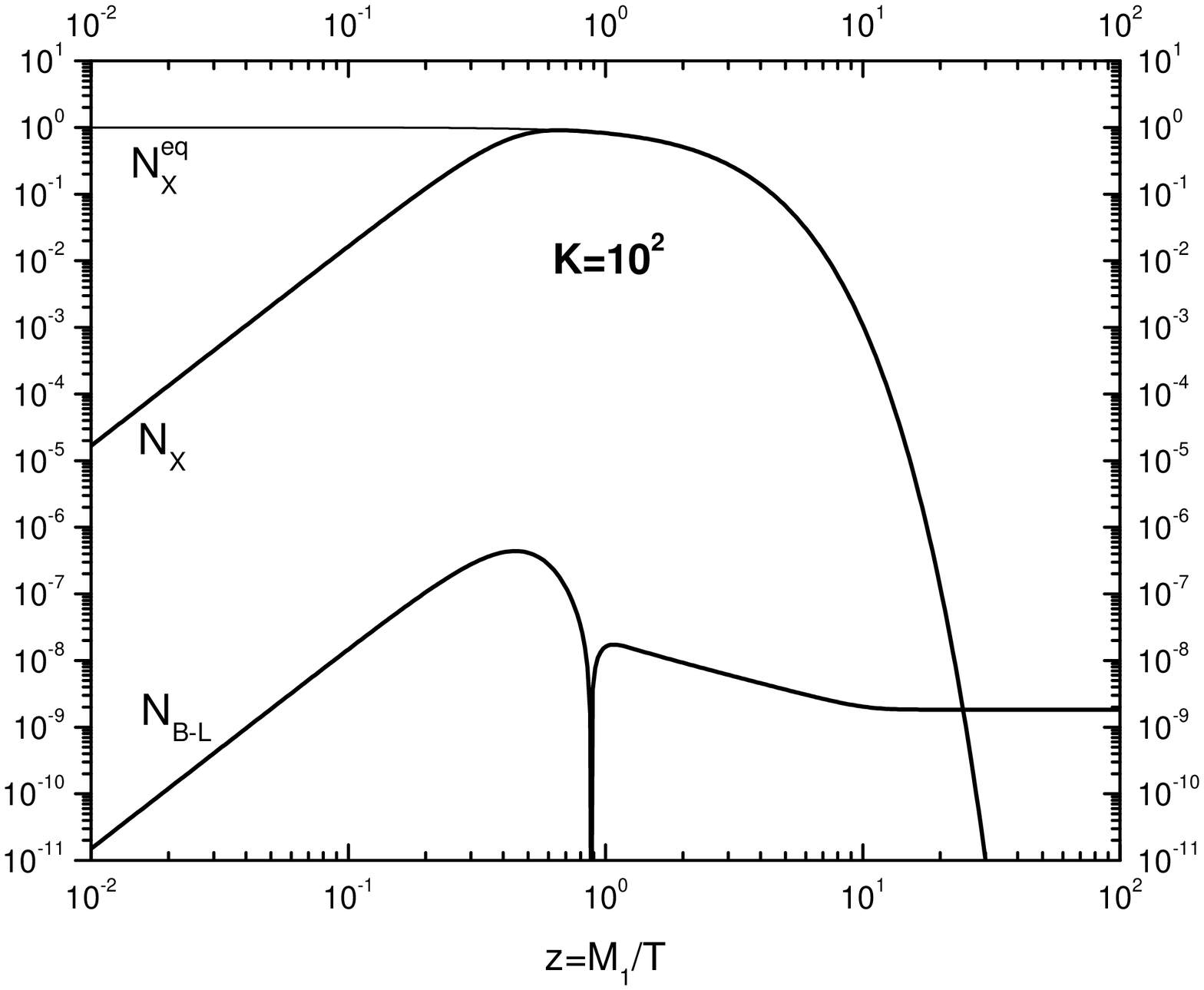,height=9cm,width=12cm}}
\vspace{-1mm}
\caption{\small fast thermalization of the X abundance in the strong wash-out regime.
The final $N_{B-L}$ abundance (for $\ve_1=0.75\times 10^{-6}$) is the same
as in the case of an initial thermal abundance (cf. Fig. 3) and it is
independent on the evolution at $z\ll z_B$.}
\label{strong}
\end{figure}
In Fig. 5 we show how even starting from a zero abundance,
the $X$'s are rapidly produced by inverse decays in a way that well before
$z_B$ the number of decaying neutrinos is always equal to the thermal number \cite{Fry}.
The final asymmetry does not even depend on the initial temperature
as far as this is higher than $\sim T_B$ and thus if one relaxes
the assumption $z_i\ll 1$ to $z_i\lesssim z_B-\Delta z_{B}$,
the final efficiency factor gets just slightly reduced (for example for $\Delta z_B\simeq 2$
this is reduced approximately by $10\%$).

Summarizing we can say that that {\em in the strong wash out regime the reduced
efficiency is compensated by the remarkable fact that, for $T_{\rm i}\gtrsim T_B$,
the final asymmetry does not depend on the initial
conditions and all non relativistic approximations work very well}.
These conclusions change quite drastically in the weak wash-out regime.

\subsection{Weak wash-out regime}

For $K\lesssim 1$ one can see that $z_B$ rapidly tends to unity (cf. (\ref{fit})).
In Fig. 2 the analytic solution for the efficiency factor, Eq. (\ref{kfth}),
is compared with the numerical solution.
It can appear surprising that, in the case of an initial thermal abundance,
the agreement is excellent not only at large $K\gtrsim 4$,
but also at small $K\lesssim 0.4$, with some appreciable deviation just around
$0.4\lesssim K\lesssim 4$.
The reason is that when the wash-out processes get frozen,
the efficiency factor depends only
on the initial number of neutrinos and not on its derivative and thus
the approximation Eq. (\ref{approx}) introduces a sensible
error only in the transition regime $K\sim 1$.

The Eq. (\ref{kfth}) can be easily generalized to any value of the initial abundance
until one can neglect the $X$'s produced by inverse decays.
More generally, one has to calculate such a contribution and it is convenient
to consider the limit case of a  zero initial $X$ abundance.
The $X$ production lasts until $z=z_{\rm eq}$, when the abundance is equal to the
equilibrium value, such that
\be
N_{X}(z_{\rm eq})=N_{X}^{\rm eq}(z_{\rm eq})  \, .
\ee
At this time the number of decays equals the number of inverse decays.
For $z\leq z_{\rm eq}$ decays can be neglected and the Eq. (\ref{ek1}) becomes
\be
{dN_X \over dz}=D(z)\,N_{X}^{\rm eq}(z)  \,.
\ee
For $z\ll 1$ one then simply finds
\be
N_{X}(z)={K\over 6}\,z^3  \, .
\ee
In the weak wash out regime the equilibrium is reached very late,
when neutrinos are already non relativistic and $z_{\rm eq}\gg 1$.
In this way one can see that the number of $N_X$ reaches,
at $z\simeq z_{\rm eq}$, a maximum value given by
\be
N_{X}(z_{\rm eq})\simeq N(K)\equiv {3\,\pi\over 4}\,K \, .
\ee
It is possible to interpolate
between the two asymptotical regimes getting a global solution for
any $z\leq z_{\rm eq}$. For $z>z_{\rm eq}$ inverse decays can be neglected
and the $X$'s decay out of equilibrium in a way that
\be
N_{X}(z>z_{\rm eq})\simeq N_X(z_{\rm eq})\,e^{-\int_{z_{\rm eq}}^z\,dz'\,D(z')} \, .
\ee
Let us now consider the evolution of the asymmetry calculating the
efficiency factor. Its value can be conveniently decomposed as the sum of two
contributions, a negative one, $\kappa^-_{\rm f}$, generated at $z<z_{\rm eq}$,
and a positive one, $\kappa^+_{\rm f}$, generated at $z>z_{\rm eq}$.
In the limit of zero wash-out we know that the final efficiency factor
must vanish, since we are assuming an initial zero abundance. This implies
that the negative and the positive contributions have to cancel each other.
The effect of wash-out is to suppress the negative contribution more than
the positive one, in a way that the cancellation is only partial.
In the weak wash-out regime it is possible in first approximation
to neglect completely the wash-out at $z\geq z_{\rm eq}$.  In this way
it is easy to derive from the Eq. (\ref{kf}) the following expression
for the final efficiency factor:
\be\label{kfw}
\kappa_{\rm f}\simeq N(K)-2\left(1-e^{-{1\over 2}\,N(K)}\right) \, .
\ee
One can see how it vanishes at the first order in $N(K)\propto K$
and only at the second order one gets $k_f\simeq (9\,\pi^2/64)\,K^2$ \cite{Fry}.

\subsection{Final efficiency factor: summary}

Generalizing the procedure seen for the strong wash-out it is possible to
find a global solution for $\kappa_{\rm f}(K)$ valid for any $K$. The calculation
proceeds separately for $\kappa^-$ and $\kappa^+$ and the final results are given by
\be\label{kf-}
\kappa^{-}_{\rm f}(K) = -2\ e^{-{1\over 2}\,N(K)}
\left(e^{{1\over 2} \overline{N}(K)} - 1 \right)
\ee
and
\be\label{kf+}
\kappa^{+}_{\rm f}(K)={2\over z_B(K)\,m\,K}
\left(1-e^{-{1\over 2} z_B(K)\,m\,K \overline{N}(K)}\right) \; .
\ee
The function $\overline{N}(K)$ extends, approximately, the definition of $N(K)$
to any value of $K$
\be
\overline{N}(K) = {N(K)\over\left(1 + \sqrt{{N(K)\over N_{\rm eq}}}\right)^2}\; .
\ee
The sum of the Eq.'s (\ref{kf+}) and (\ref{kf-}) is plotted, for $m=1$, in Fig. 2 (short-dashed line)
and compared with the numerical solution (solid line).

We can now outline some conclusions about a comparison between the weak
and the strong wash-out regimes. A large efficiency in the weak wash-out
regime relies on some unspecified mechanism that should have produced a large
(thermal or non thermal) $X$ abundance before their decays.
On the other hand the decrease of the
efficiency at large $K$ in the strong wash-out regime is only (approximately) linear
and not exponential \cite{kt}. This means that for moderately large values of $K$
a small loss in the efficiency would be compensated by a
full thermal description such that the predicted asymmetry
does not depend on the initial conditions, a nice situation that
resembles closely the Standard Big Bang Nucleosynthesis
scenario for the calculation of the primordial nuclear abundances.

\section{Leptogenesis}

Let us see now how the results that hold for generical
{\em baryogenesis models from heavy particle decays} get specialized  in the
case of leptogenesis \cite{fy}.  This is the cosmological
consequence of the seesaw mechanism, explaining
the lightness of the ordinary neutrinos through the
existence of three new heavy RH neutrinos
$N_1,N_2,N_3$ with masses respectively
$M_1\leq M_2 \leq M_3$, much larger than the electroweak scale.
The simple seesaw formula,
\begin{equation}\label{seesaw}
m_\n = - m_D {1\over M} m_D^T\; ,
\end{equation}
relates the neutrino mixing matrix $m_{\nu}$ to the
RH neutrino mass matrix $M$ and to the Dirac neutrino mass
matrix $m_D=h\,v$ generated by the Yukawa coupling matrix $h$, where $v$
is the Higgs vacuum expectation value.
Both light and heavy neutrinos are predicted to be Majorana neutrinos.
All mass matrices are in general complex and this provides
a natural source for the $C\!P$ asymmetry while the new RH neutrinos are
the natural candidates to play the role of the $X$ particles.
In this case things are apparently more complicated since there are
three of them.  We will assume that the decays and inverse decays
of the two heavier neutrino decays do not
influence the value of the final asymmetry. This assumption holds for example either if
the asymmetry produced by the two heavier RH neutrinos is negligible or
if this is produced and then washed out by the inverse decays of the lightest
(heavy RH). In this way we can straightforwardly apply the general picture of
baryogenesis from $X$ decays to leptogenesis,
with the $N_1$'s playing the role of the $X$ particles.

\subsection{Decay parameter and neutrino masses}

The total $N_1$ decay width is given by
\be\label{Gamma}
\Gamma_D^{\rm rest}= {\widetilde{m}_1\,M_1^2\over 8\,\pi\,v^2} \, ,
\ee
where the {\em effective neutrino mass} is defined as \cite{plum}
\be\label{mt}
\widetilde{m}_1={(m_D^{\dagger}\,m_D)_{11}\over M_1} \, .
\ee
It is then easy to see that the decay parameter is
related to $\widetilde{m}_1$ \cite{yasu} by the following relation
\be\label{Klep}
K={\widetilde{m}_1\over m_{\star}} \, ,
\ee
where the equilibrium neutrino mass $m_{\star}$ can
be written as
\be\label{mstar}
m_{\star}={v^2\over M_{\star}}\simeq 10^{-3}\,{\rm eV} \, ,
\ee
with the quantity $M_{\star}$ given by
\be
M_{\star}={3\,\sqrt{5}\over 16\,\pi^{5/2}}\,{M_{Pl}\over\sqrt{g_{\star}}}
\simeq 3\times 10^{16}\,{\rm GeV} \, .
\ee
It is quite non trivial that the value of $m_{\star}$ is close
to the neutrino mixing mass scales and we will show soon the relevance
of this result. For the moment note that the value of $m_{\star}$
is independent on the well known success of the seesaw mechanism
in explaining the atmospheric and solar neutrino mass scales and this is
why we wrote $m_{\star}$ in a sort of seesaw-like form, introducing
the scale $M_{\star}$. Apart from the very generical consideration that
the logarithm of $M_{\star}$ is expected to be close to the Planck scale,
this is not related to the grand unified scale, rather to the expansion rate
at the baryogenesis time
\footnote{It is then quite curious that the value
of $M_{\star}$ is just the value of the supersymmetric unification scale.}.


Let us now assume that the simple decays plus inverse decays picture
studied in the previous section is a good approximation of leptogenesis.
It is then crucial to determine the value of the the effective neutrino
mass $\widetilde{m}_1$, and thus, from the Eq. (\ref{Klep}), the value
of the decay parameter $K$, in order to answer the important question whether
leptogenesis lies in the strong or in the weak wash-out regime.

It is always possible to work in a basis in which the heavy neutrino mass matrix
is diagonal, such that $M={\rm diag}(M_1,M_2,M_3)\equiv D_M$.
Moreover one can also simultaneously diagonalize the light neutrino
mass matrix $m_\n$ by mean of the unitary MNS matrix $U$, such that
\be
U^{\dagger}\,m_\n\,U^{\star}=-D_m \, .
\ee
In this way the seesaw formula (\ref{seesaw}) gets specialized
in the following way:
\be
D_m=U^{\dagger}\,m_D\,D_M^{-1}\,m_D^{T}\,U^{\star} \, .
\ee
This expression can be also re-casted as an orthogonality condition,
\be
\Omega\,\Omega^T=\Omega^T\,\Omega=I \, ,
\ee
for the $\Omega$ matrix defined as \cite{casas}
\be\label{Omega}
\Omega=D_m^{-1/2}\,U^{\dagger}\,m_D\,D_M^{-1/2}
\ee
and whose matrix elements are then simply given by
\be\label{Omegah}
\Omega_{ij}={v\,\,\widetilde{h}_{ij}\over \sqrt{m_i\,M_j}} \, ,
\ee
where $\widetilde{h}=U^{\dagger}\,h$. The $\Omega$ matrix
is fully determined by three complex parameters. Four of them are needed
to fix the three first column entries $\Omega_{11}$,
$\Omega_{21}$ and $\Omega_{31}$, particularly important for leptogenesis.
This because if one inverts the relation (\ref{Omega}), in a way to get an expression
of $m_D$ in terms of $\Omega$, and plugs it into
the effective neutrino mass definition (cf. (\ref{mt})), then one easily gets \cite{hama}
\be\label{mt1}
\mt=m_1\,|\O_{11}^2|+m_2\,|\O_{21}^2|+m_3\,|\O_{31}^2|  \, .
\ee
From the orthogonality of $\Omega$ it follows that $\mt\geq m_1$. This
is the only fully model independent restriction on $\mt$. For configurations such that
\be
\sum_j\,|\O_{j1}^2| \sim \left|\sum_j\,\Omega_{j1}^2\right|=1
\ee
one has $\mt\lesssim m_3$. Models with $\mt\gg m_3$ rely on
the possibility of strong phase cancellations.

Neutrino mixing data provide two important pieces of information on the neutrino mass spectrum.
In the case of normal hierarchy one has $m^2_3-m^2_2=\Delta m^2_{\rm atm}$ and
$m^2_{2}-m^2_1=\Delta m^2_{\rm sol}$. In the case of inverted hierarchy
$m^2_3-m^2_2=\Delta m^2_{\rm sol}$ and $m^2_2-m^2_1=\Delta m^2_{\rm atm}$.
 The third, still undetermined, independent information, the absolute neutrino mass scale,
can be conveniently expressed in terms of the lightest neutrino mass $m_1$.
 The two heavier neutrino masses are then given, for normal (inverted) hierarchy, by
\begin{eqnarray}
\label{numanor1}
m_3^{\,2}
&=& m_1^2 + m_{\rm atm}^2\;, \\
\label{numanor2}
m_2^{\,2} &=& m_1^2 + \Delta m^2_{\rm sol}\,(\Delta m^2_{\rm atm}),
\end{eqnarray}
where we defined $m_{\rm atm}=\sqrt{\Delta m^2_{\rm atm} + \Delta m^2_{\rm sol}}$.
The latest measurements give \cite{atm}
\begin{equation}
\D m^2_{\rm atm}= (2.6\pm 0.4)\times 10^{-3}\,{\rm eV}^2\;,
\end{equation}
and for solar neutrinos \cite{solar}
\begin{equation}
\D m^2_{\rm sol}\simeq  (7.1^{+1.2}_{-0.6} \times 10^{-5})\,{\rm
eV^2}\;,
\end{equation}
from which it follows that
\begin{equation}\label{matm}
m_{\rm atm}=(0.051\pm 0.004)\,{\rm eV} \, .
\end{equation}
These relations imply that for $m_1^2\gg m_{\rm atm}^2$ neutrinos are
quasi-degenerate ($m_3\simeq m_2\simeq m_1$), whereas for
$m_1^2\ll m_{\rm atm}^2$ they are hierarchical ($m_1\ll m_2,m_3$).

For fully hierarchical neutrinos ($m_1=0$) there is practically no
restriction on $\mt$. However the case $\mt\ll m_2,m_3$ requires
$|\O_{21}^2| << 1$ and $|\O_{31}^2|<< m_2/m_3$.
This situation cannot be excluded \cite{bdp1} but,
because of the observed large mixing angles in the mixing matrix $U$,
it relies on a fine tuning between the $U$ and $m_D$
matrix elements (cf. (\ref{Omega})), such that the off-diagonal terms
are very small. This qualitative and general
argument is supported by different investigations on specific models or classes of models
for which typically one finds
$m_{\rm sol}\simeq m_2\lesssim \mt \lesssim m_3\simeq m_{\rm atm}$ \cite{ir}.
Therefore, in the case of normal hierarchy
one has that the favored range for the $\mt$ value
is given by ${\cal O}(m_{\rm sol})\leq \mt \leq {\cal O}(m_{\rm atm})$, that
in terms of the decay parameter (cf. (\ref{Klep})) gets translated into the range
\be\label{Kleprange}
{\cal O}(K_{\rm sol}\simeq 7)< K < {\cal O}(K_{\rm atm}\simeq 50) \, ,
\ee
while for inverted hierarchy the situation is even simpler since
$\mt={\cal O}(m_{\rm atm})$ and $K={\cal O}(K_{\rm atm}\simeq 50)$.
One thus arrives to the interesting conclusion that
neutrino mixing data favor leptogenesis to lie in a {\em mildly strong wash out regime},
strong enough to benefit from the advantages we discussed, independence on the initial conditions
plus minimal theoretical uncertainties, but not too much to result in an untenable
efficiency loss. This conclusion derives because both the
two independent experimental quantities, $m_{\rm sol}$ and $m_{\rm atm}$, are about
ten times $m_{\star}$ and so now one can better appreciate the nice matching
of the theoretical quantity $m_{\star}$ with the experimental data
\footnote{Note that this is also a consequence of  the recent exclusion of the low solution
 in the solar neutrino data, that would have implied $K_{\rm sol}\ll 1$.}.
In the range $K_{\rm sol}\lesssim K \lesssim K_{\rm atm}$ a good fit of the
 final efficiency factor (cf. Eq. (\ref{kfth})) is given by the power law
\be\label{kfpllep}
\k_{\rm f}={0.5\over K^{1.2}} \simeq 3\times 10^{-2}\,
\left({10^{-2}\,{\rm eV}\over \mt} \right)^{1.2} \, ,
\ee
shown in Fig. 2 (dot-dashed line).
These conclusions hold under the assumption that leptogenesis is well
approximated by the simple decays plus inverse decays picture and we have now
to verify whether they are drastically modified or just corrected by the
account of $N_1$ scatterings and $\Delta L=2$ processes.

\subsection{Scatterings}

The $N_1$'s can also be destroyed or produced in $\Delta L=1$
scatterings involving the top quark. These are mediated by the Higgs and
can occur in the s channel, like $N_1+l \leftrightarrow t+q$, or in the
t channel, like $N_1+t \leftrightarrow l+q$.
The account of these processes modify the kinetic equations
(\ref{ek1}) and (\ref{ek2}) in the following way:
\bea\label{ekS1}
{dN_{N_1}\over dz}& = & -(D+S)\,(N_{N_1}-N_{N_1}^{\rm eq}) \, , \\  \label{ekS2}
{dN_{B-L}\over dz}& = & -\ve_1\,\;D\,(N_{N_1}-N_{N_1}^{\rm eq})-
(W_{ID}+W_{\Delta L+1})\,N_{B-L} \, .
\eea
Note that scatterings have two effects: they contribute
both to the  neutrino production (the $S$ function) and
to the wash-out (the $W_{\Delta L+1}$ function).

The first one is important in the weak wash-out regime. As one can see from the
Eq.'s (\ref{ekS1}) and (\ref{ekS2}), the production of the $N_1$'s from the $S$
function is not associated to a production of the asymmetry, simply because these
processes do not violate $C\!P$. In Fig. 5 (from \cite{pedestrians}) we show an example
of $N_1$ production for
\begin{figure}[t]
\centerline{\psfig{file=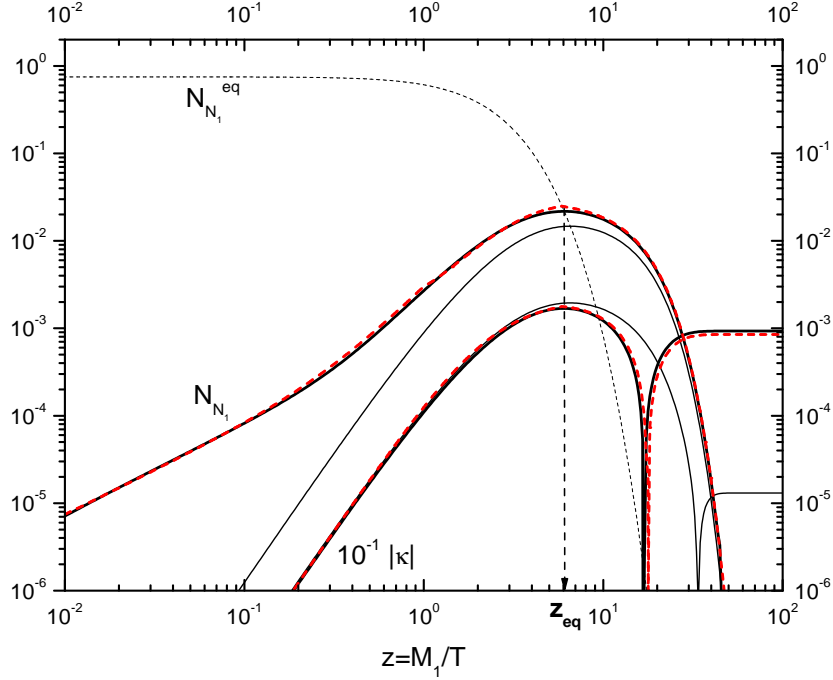,height=11cm,width=14cm}}
\vspace{-5mm}
\caption{\small comparison between the case when scatterings are included (thick lines)
with the decays plus inverse decays picture (solid thin lines) for $\mt=10^{-5}\,{\rm eV}$.}
\label{Sprod}
\end{figure}
$\mt=10^{-5}\,{\rm eV}$ ($K\simeq 0.01$), comparing the case when scatterings are included
with the case when they are neglected.
It can be seen how at $z=z_{\rm eq}$ the number of neutrinos is approximately doubled
while the final asymmetry is two orders of magnitude larger. The reason is that the
neutrino production from the scatterings is not associated to a production of a negative
asymmetry. On the other hand all produced neutrinos yield a positive
contribution when they decay.
The expression (\ref{kfw}) for the final efficiency factor in the weak wash-out regime
gets thus modified in the following way at the first order in K
\be\label{kfwS}
\kappa_{\rm f}\simeq
\left.{N_{N_1}\,D\over D+S}\right|_{z=z_{\rm eq}}-N(K) \propto K \, .
\ee
If scatterings are switched off the negative and the positive contribution
cancel at the first order like we saw already. If $S\neq 0$ the positive term is enhanced
while the negative one remains unchanged and in this way the sum does not vanish any more.
Hence this effect makes more efficient the asymmetry production at small $K$, without
having to assume an initial thermal abundance. There is however a drawback. The final result
is quite sensitive to the theoretical assumptions. The scattering cross section depends
on the ratio, $M_h/M_1$, of the Higgs mass to the RH neutrino mass. The case depicted
in Fig. 5 is for $M_h/M_1=10^{-5}$. For smaller values of this ratio the result
does not change much. However it has been recently pointed out \cite{gnrrs}
that the Higgs mass is better described by its thermal mass such that
$M_h/M_1 \simeq 0.4/z$.
 The relevant values of $z$ for neutrino production are $z\simeq 1$ and so the
ratio $M_h/M_1 \simeq 0.4$. Such an high value has the effect to suppress heavily
the $S$ term and the suppression is made even stronger by the account of the running
of the top Yukawa coupling at high temperature.
In this way the simple decays plus inverse decays picture is practically recovered.
\begin{figure}[t]
\centerline{\psfig{file=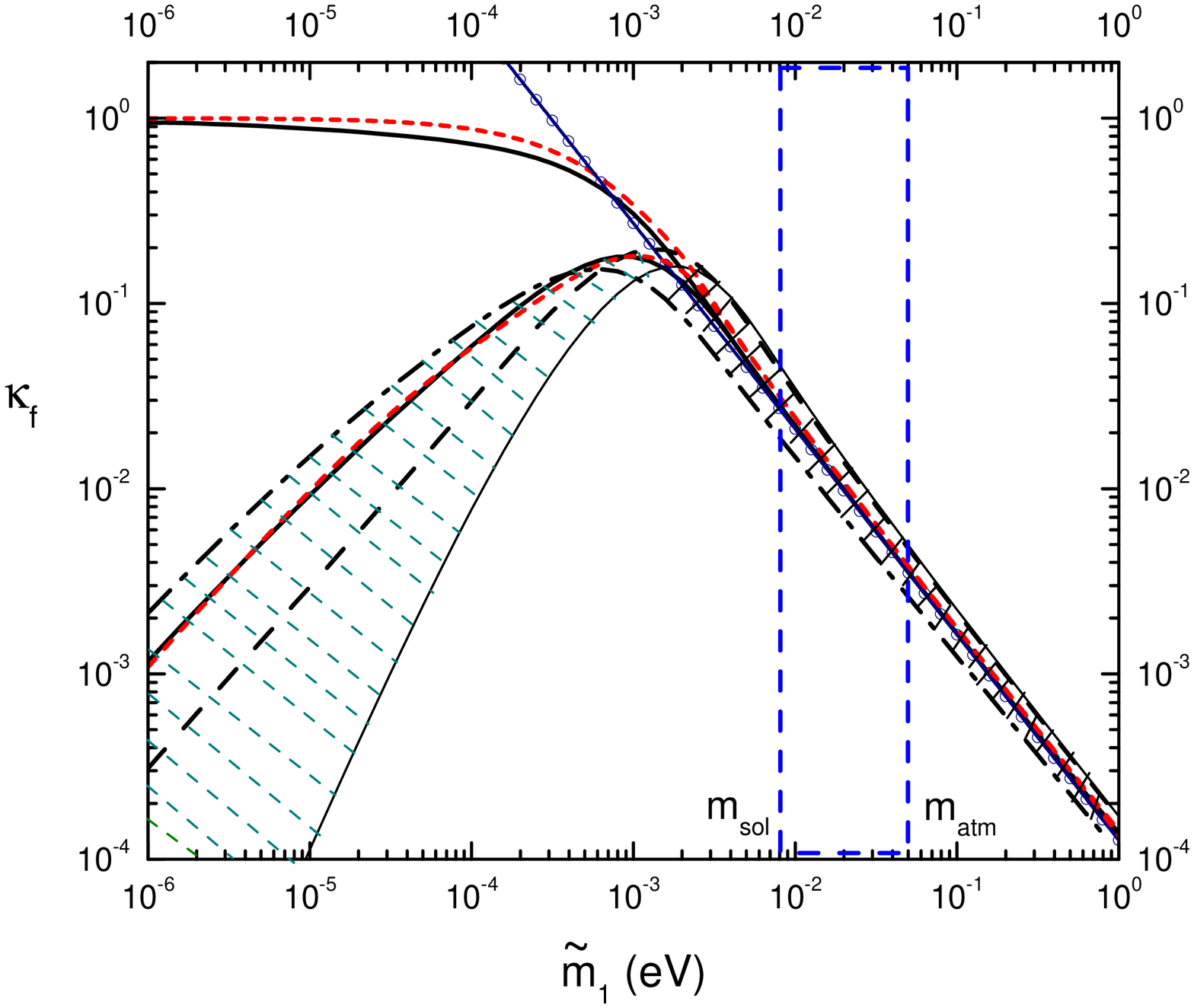,height=9cm,width=14cm}}
\vspace{-5mm}
\caption{\small efficiency factor when scatterings are included.}
\label{Skf}
\end{figure}
On the other hand in \cite{pilun,gnrrs} it has been noticed how scatterings
involving gauge bosons should also be included. These scatterings yield an
additional contribution to the S function such that the final result is between  a situation where
scatterings are neglected and one where scatterings involving top quark and small
$M_h/M_1$ are taken into account.  The conclusion is that in the weak-wash out regime
the theoretical uncertainties are such that it seems that any result between the simple decays
plus inverse decays picture, for which $\k_{\rm f}\propto K^2$ or a behavior
$\k_{\rm f}\propto K$ (cf. Eq. (\ref{kfwS}) ) cannot be firmly excluded at the moment.
These large theoretical uncertainties, represented in Fig. 6 with the short-dashed region,
are in addition to the model dependence in the description of the initial conditions.

In the {\em strong wash-out  regime} all difficulties get considerably reduced.
The theoretical uncertainties in the description od scatterings can change the final
efficiency factor no more than $(20\div 30)\%$ and this is clearly shown in Fig. 6,
where at large $K$ the (thin solid line) range shrinks considerably
compared to the (short-dashed line) range at small $K$. This because
the thermal abundance limit is saturated at $z_{\rm eq}\ll 1$ anyway
and therefore the number of decaying neutrinos does not depend on the $S$ function. A residual
source of uncertainty is still present because of the scattering contribution, $W_{\Delta L=1}$,
to the wash-out.
The effect of this term is however small, for the simple reason that in the strong wash out regime
the surviving asymmetry is produced sharply around $z_{B}\gg 1$ and at such low temperatures
inverse decays are dominant compared to scatterings.
The conclusion is that in the strong wash-out regime the simple decays plus inverse
decays picture does not get modified by scatterings within the
theoretical uncertainties. It has been also pointed out \cite{spectator} that an
accurate description of the dynamics of sphalerons in converting the
lepton number into a baryon number is expected to lead to a suppression of the final
asymmetry of a ${\cal O}(1)$ factor and since this is currently neglected it gives
an additional contribution to the theoretical uncertainties.
Taking into account all these effects, an expression for the
final efficiency factor in the strong wash out regime that  accounts
for the theoretical uncertainties is given by the power law \cite{pedestrians}
\be\label{plaw}
\k_{\rm f}=(2\pm 1)\,10^{-2}\,\left({10^{-2}\,{\rm eV}\over\mt}\right)^{1.1\pm 0.1}  \, .
\ee
The central value corresponds to the curve represented  in Fig. 5 with circles
(more precisely this is obtained for a power law $\mt^{-1.13}$),
while the range that is spanned by the error, corresponds approximately
to the thin solid line area. The upper values of this range is the power law
(\ref{kfpllep}), that well describes the simple decays plus inverse decays
picture where the wash-out from scatterings is neglected
\footnote{The result obtained in \cite{gnrrs} corresponds to this situation because
at $z\simeq z_B\gg 1$ the Higgs thermal mass suppresses the wash-out from scatterings
involving the top quark while the contribution from scatterings involving gauge bosons
is negligible.}.

\subsection{$\Delta L=2$ processes}

There is another important contribution to the wash-out term arising from
the $\Delta L=2$ processes like $lH\leftrightarrow \bar{l}\bar{H}$ and mediated by the RH
neutrinos. In the non relativistic regime this contribution tends simply to  \cite{fy2}
\be\label{nonr}
\Delta W (z\gg 1)\simeq {\o\over z^2} \left({M_1\over 10^{10}\,{\rm GeV}}\right)
\left({\mb\over {\rm eV}}\right)^2 \;,
\ee
with $\mb^2=m_1^2+m_2^2+m_3^3$, and dominates on the other
Boltzmann suppressed wash-out terms arising from inverse decays and scatterings.
A well known problem is that at temperatures $T \sim M_1$ one has to be sure that the cross section
of $\Delta L=2$ processes does not double count the on-shell contribution already
accounted by inverse decays followed by decays to a final state with opposite lepton number
(i.e. $l+\bar{H}\rightarrow N_i \rightarrow \bar{l}+H$).
In \cite{bdp1} it has been found that the subtraction procedure
usually employed in the previous literature
gives arise to a washout $\Delta L=2$ term that is very well
approximated by the asymptotical non-relativistic limit Eq. (\ref{nonr})
plus a term that is just half the washout from inverse decays. In \cite{gnrrs}
this second term has been shown to be spurious and to disappear when a proper subtraction procedure
is employed. This result has been confirmed in \cite{pedestrians}.
Therefore the effect of the $\Delta L=2$ processes is entirely
well approximated by its non-relativistic limit.
It is easy to see that for $M_1\ll 10^{14}\,{\rm GeV}\,(0.05\,{\rm eV}/\mb)^2$,
this term can be neglected. Thus, for sufficiently small neutrino masses and
in the strong wash-out regime, we can conclude that leptogenesis is well approximated
by a simple decays plus inverse decays picture.

\subsection{$C\!P$ asymmetry and seesaw geometry}

So far we concentrated on the {\em kinetic theory of leptogenesis}
and we have seen how neutrino mixing data favor a very simple regime
in which predictions are model independent and theoretical uncertainties
are minimized. We have now to answer the crucial question
whether the resulting final asymmetry can explain the
measured CMB value (cf. (\ref{etaBCMB})).
 The thermodynamical point of view, i.e. the efficiency factor,
 is not enough to answer this question, since one needs to know
the value of the $C\!P$ asymmetry too. This is a specific
leptogenesis issue that concerns what can be called the {\em seesaw geometry}.

 A perturbative calculation from the interference between tree level
 and vertex plus self energy one-loop diagrams yields \cite{CPas}
\be\label{eps1g}
\varepsilon_1 \simeq  {1\over 8\pi}
\sum_{i=2,3}\,{{\rm Im}\,\left[(h\,h^{\dagger})^2_{i1}\right]\over (h\,h^{\dagger})_{11}} \,\times\,
\left[f_V\left({M^2_i\over M^2_1}\right)+f_S\left({M^2_i\over M^2_1}\right)\right] \, .
\ee
The function $f_V$, describing the vertex contribution, is given by
\be
f_V(x)=\sqrt{x}\,\left[1-(1+x)\,\ln\left({1+x\over x}\right)\right] \, ,
\ee
while the function $f_S$, describing the self-energy contribution, is given by
\be
f_S(x)={\sqrt{x}\over 1-x}  \, .
\ee
In the limit $x\gg 1$, corresponding to have a mild RH neutrinos mass hierarchy
with $M^2_{2,3} \gg M^2_1$, one has
\be\label{flimit}
f_V(x)+f_S(x)\simeq -{3\over 2\,\sqrt{x}} \, .
\ee
In this limit and barring strong phase cancellations \cite{hambye} the
expression (\ref{eps1g}) simplifies into \cite{fred}
\be\label{eps1h}
\varepsilon_1 \simeq  -{3\over 16\pi}
\,{{\rm Im}\,\left[(h^{\dagger}\,h\,M^{-1}\,h^T\,h^{\star})_{11}\right]
\over (h\,h^{\dagger})_{11}} \, .
\ee
Replacing $h$ with $\Omega$ (cf. (\ref{Omegah}) ) one then gets \cite{di}
\be\label{eps1Om}
\varepsilon_1 \simeq  -{3\over 16\pi}\,{M_1\,m_{\rm atm}\over v^2}\,
\beta(m_1,\mt,\Omega^2_{j1}) \, ,
\ee
where we introduced the convenient quantity
\be\label{beta}
\beta(m_1,\mt,\Omega^2_{j1})=
{\sum_j\,m^2_j\,{\rm Im}(\Omega^2_{j1})\over
m_{\rm atm}\,\sum_j\,m_j\,|\Omega_{j1}^2|}  \, .
\ee
The final asymmetry is proportional to the product of the $C\!P$ asymmetry
times the final efficiency factor that, in the simplified
decays plus inverse decays picture, depends only on the effective neutrino mass
$\widetilde{m}_1$.
The expression (\ref{eps1Om}) shows that the $C\!P$ asymmetry depends
on the three complex numbers $\Omega^2_{j1}$ and thus
it introduces a model dependence in the prediction of the final asymmetry
that one was hoping to have removed in the calculation of the final efficiency factor.
It is however possible to maximize the absolute value of the $C\!P$ asymmetry
respect to the `geometrical' parameters $\O_{j1}^2$, thus finding a non trivial maximum
$\ve_1^{\rm max}(M_1,\mt,m_1)$ depending only on
$M_1$, $\mt$ and $m_1$. One can then define an {\em effective leptogenesis phase}
$\d_L$ such that the expression (\ref{beta}) can be re-casted in the following way
\be\label{betad}
\beta(m_1,\mt,\Omega^2_{j1})=\beta_{\rm max}(m_1,\mt)\,\sin\d_L(\mt,m_1,\Omega_{j1}^2) \, .
\ee
The maximum of the absolute value of the $C\!P$ asymmetry and of the function $\beta$
are thus realized for those particular geometrical configurations, corresponding to some
$\Omega^2_{j1}$'s values, such that $\sin\d_L=1$. A general
procedure for the calculation of $\ve_1^{\rm max}$ and $\sin\d_L$ is presented
in \cite{prep}. Here we just sketch some general features and
describe two particularly interesting limit cases.

 If one represents the three $\Omega_{j1}^2$ in the complex plane, the orthogonality
condition fixes the sum of the three to start from the origin and to
end up onto the real axis at the point Re$(\sum_j\Omega^2_{j1})=1$,
as shown in Fig. 8 for a generic configuration (solid line arrows).
\begin{figure}[t]
\centerline{\psfig{file=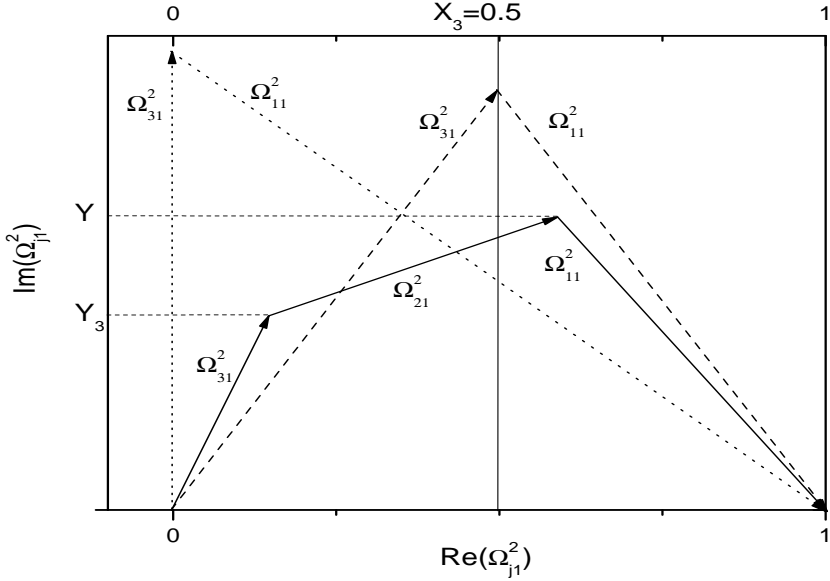,height=9cm,width=14cm}}
\vspace{-5mm}
\caption{\small Seesaw geometry. Different configurations of $\Omega^2_{j1}$
(see text for explanation) with the same value of $\mt$.}
\label{kfS}
\end{figure}
 Using the orthogonality condition, defining $\Omega_{j1}^2=X_j+i\,Y_j$
and using the definition of $\mt$ (cf. \ref{mt1}), this can be re-casted as
\be
\beta(m_1,\mt,\Omega^2_{j1})={\Delta m^2_{32}\,Y_3+\Delta m^2_{21}\,Y\over m_{\rm atm}\,\mt} \, ,
\ee
with $Y\equiv Y_2+Y_3$.
The absolute value of $\beta$ has to be maximized for $\mt$ constant.
In general one always finds that \cite{di,bdp3}
\be\label{betamax}
\beta_{\rm max}(m_1,\mt)=\beta_{\rm max}(m_1)\,f(m_1,\mt) \leq 1 \, ,
\ee
with $\beta_{\rm max}(m_1)=(m_3-m_1)/m_{\rm atm}$,
$f(m_1,\mt)\leq 1$ and $f(m_1,\infty)=1$.

An interesting limit case is that of {\em fully hierarchical neutrinos}
for $m_1=0$. In this case $\beta_{\rm max}=1$ and there is no global
suppression. Moreover one has $\mt=m_2\,|\Omega_{21}^2|+m_3\,|\Omega_{31}^2|$
and thus, for any change configuration such that $|\Omega_{21}^2|$
and $|\Omega_{31}^2|$ are constant, the quantity $\mt$ is also constant
while  $|\Omega_{11}^2|$ can be arbitrarily modified. Hence
$|\beta(\mt,m_1,\Omega_{j1}^2)|$ is maximized for
configurations such that $X_2=X_3=0$. It is then easy to see that it
is further maximized for $Y_2=0$ and $Y_3=\mt/m_3$, corresponding to the
configuration shown in Fig.8 with dotted line arrows. In this case one has
very simply $f(0,\mt)=1$. Therefore, the case $m_1=0$
corresponds, for a fixed $M_1$, to an absolute maximum of the $C\!P$ asymmetry
given by \cite{asaka,di} (cf. (\ref{eps1Om}) and (\ref{beta}))
\be\label{e1maxM1}
\ve_1^{\rm max}(M_1)=
{3\over 16\pi}\,{M_1\,m_{\rm atm}\over v^2}
\simeq 10^{-6}\,\left({M_1\over 10^{10}\,{\rm GeV}}\right)
\,\left({m_{\rm atm}\over 0.05\,{\rm eV}}\right) \, .
\ee
Note that with this last definition of $\ve_1^{\rm max}(M_1)$,
together with the expressions (\ref{betad}) and (\ref{betamax}),
the Eq. (\ref{eps1Om}) for the $C\!P$ asymmetry can be re-casted like
\be
\ve_1=-\ve_1^{\rm max}(M_1)\,\b_{\rm max}(m_1)\,f(m_1,\mt)\,\sin\d_L(m_1,\mt,\O_{j1}^2) \, ,
\ee
showing the sequence of different maximization steps.

In the {\em quasi-degenerate limit}
the expression (\ref{mt1}) for $\mt$ becomes simply $\mt\simeq m_1\sum_j\,|\Omega_{j1}^2|$.
Thus  the condition $\mt={\rm const}$ is equivalent to select all those configurations
for which $\sum_j\,|\Omega_{j1}^2|$ is constant.
Hence it is straightforward to conclude that $|\beta|$ is maximum for a configuration
such that $Y_2=X_2=0$ and $X_3=1/2$, shown in Fig. 8 with dashed line arrows.
 Using that in the quasi degenerate limit
$m^2_{\rm atm}\simeq 2\,m_1\,(m_3-m_1)$, one obtains
\footnote{This limit expression has been first shown in
\cite{hambye} using the approximation $m_{\rm sol}=0$. Here we derived it
in a more general way.}
\be
f(m_1,\mt)=\sqrt{1-{m_1^2\over\mt^2}}  \, .
\ee
Note that in both the two limit cases the maximum $C\!P$ asymmetry is obtained for configurations
such that $\O_{21}=0$.
It is possible to show that this result holds in general,
for any value of $m_1$ \cite{prep}.

Therefore, for maximal $C\!P$ asymmetry ($\sin\delta_L=1$), one can still express
all predictions in terms just of $M_1,\mt$ and $m_1$.
In particular it is possible to express the CMB constraints, for
all neutrino models, just in terms of these three parameters. For
specific models it can happen of course that
$\sin\d_L<1$ and the constraints get, in general, more restrictive.

\subsection{CMB bound}

From the Eq.'s (\ref{etaB}) and (\ref{NBmLf}) one obtains
for the predicted baryon to photon number ratio
\be
\eta_B =d\,\ve_1\,\k_{\rm f}\, ,
\ee
where the quantity $d$ is defined as
\be
d={a_{\rm sph}\over N_{\gamma}^{\rm rec}}  \, .
\ee
In the Standard Model case one has
$a_{\rm sph}=28/79$, while the number of photons at recombination,
assuming a standard thermal history, is given by
\be
N_{\gamma}^{\rm rec}={4\,g_{\rm SM}\over 3\,g_{\rm rec}}=
{4697\over 129} \simeq 36
\ee
and thus $d\simeq 0.97\times 10^{-2}$.

The {\em maximum baryon asymmetry} $\eta_B^{\rm max}(M_1,\mt,m_1)$
is defined like the asymmetry corresponding to the
maximum $C\!P$ asymmetry
\be
\ve_1^{\rm max}(M_1,\mt,m_1)=\ve_1^{\rm max}(M_1)\,\b_{\rm max}(m_1)\,f(m_1,\mt) \, .
\ee
The CMB bound is then simply equivalent to require
\be\label{CMBc}
\eta_B^{\rm max}(M_1,\mt,m_1) \geq \eta_B^{CMB}
\ee
and therefore will yield constraints on the space of
the three parameters $M_1$, $m_1$ and $\mt$.

\subsection{Lower bounds on the lightest RH neutrino mass and on the reheating temperature}

We have seen that the absolute maximum of the $C\!P$ asymmetry is obtained
for $m_1=0$. For $m_1>0$ the function
$\b_{\rm max}(m_1)$ suppresses the $C\!P$ asymmetry \cite{di}.
Furthermore the $\Delta L=2$ wash-out term gets enhanced
when the absolute neutrino mass scale increases (cf. (\ref{nonr})).
Therefore, the maximum baryon asymmetry
$\eta_B^{\rm max}$ is maximal when $m_1=0$.
In this case the allowed
region in the space of the parameters $M_1$ and $\mt$ and compatible with the
CMB constraint is maximum \cite{bdp1} and one finds an interesting lower bound
on the $M_1$ value \cite{di,bdp1} just plugging the expression (\ref{e1maxM1})
into the CMB constraint (cf. (\ref{CMBc}))
\bea\label{lbM1} M_1 \geq M_1^{\rm min} &=& {1\over d}\,
{16\,\pi\over 3}\,{v^2\over m_{\rm atm}}\,
{\eta_B^{CMB}\over\k_{\rm f}} \NO\\
&\simeq& 6.4\times 10^{8}\,{\rm GeV} \left(\eta_B^{CMB}\over
6\times 10^{-10}\right)
\left(0.05\,{\rm eV}\over m_{\rm atm}\right)\,\k_{\rm f}^{-1} \;.
\eea
For an {\em initial thermal abundance} and in the limit $\mt/m_{\star}\rightarrow 0$,
one has, by definition, $\k_f=1$ and so one finds
\be\label{M1lbza}
M_1 \geq
(6.6\pm 0.8)\times 10^8\,{\rm GeV}
\gtrsim 4\times 10^8\,{\rm GeV} \, ,
\ee
where the last inequality is the $3\,\sigma$ bound and we have used the
experimental values Eq. (\ref{etaBCMB}) and Eq. (\ref{matm}).
 The case of a dynamically generated $N_1$ abundance is more
 significative and in this case the peak value $\k_{\rm f}\simeq 0.18$, implies
\be\label{M1lbta}
M_1 \geq (3.6\pm 0.4) \times 10^9\,{\rm GeV}
\gtrsim 2 \times 10^9\,{\rm GeV} \, .
\ee
The most interesting situation corresponds to the
range $(m_{\rm sol}\div m_{\rm atm})$, for which the power law
Eq. (\ref{plaw}) can be used for $\k_{\rm f}$, thus giving
\be\label{M1lbsw}
M_1\gtrsim [3.3(2.2)\pm 0.4(0.3)] \times 10^{10}\,{\rm GeV}\,
\left(\widetilde{m}_1\over 10^{-2}\,{\rm eV} \right)^{1.1(1.2)}
\gtrsim [1.5(1)-10(9)]\times  10^{10}\,{\rm GeV} \, ,
\ee
where we have used the central value (upper value) in the Eq. (\ref{plaw}).
The $M_1$ lower bound can be translated
into a {\em lower bound on the initial temperature} $T_{\rm i}$ that,
within inflationary models, can be identified with the reheating temperature,
corresponding to that temperature below which a radiation dominated
regime holds \cite{gkr}. So far we assumed that this is much larger than $M_1$.
If one relaxes this assumption
then the final efficiency factor gets reduced. For small $\mt$ the threshold value
is given approximately by $M_1$ itself since below this temperature either, assuming an initial thermal
abundance, the $N_1$ abundance is thermally suppressed or the production gets considerably
suppressed for an initial zero abundance. Therefore,
for small values $\mt\lesssim 10^{-3}\,{\rm eV}$, the same bounds
(\ref{M1lbza}) and (\ref{M1lbta}) apply also to the reheating temperature.

In the more interesting case of strong wash-out, since the $90\%$ of the
surviving abundance is produced in an interval $z\simeq z_B\pm 2$, then the reheating temperature
can be $\simeq z_B-2$ times lower than $M_1$ \cite{pedestrians},
without any appreciable change in the
final predicted asymmetry. In the interesting range $(m_{\rm sol},m_{\rm atm})$
one has that $z_B$ spans between $6$ and $8$ and thus the bound Eq. ($\ref{M1lbsw}$)
gets relaxed from 4 to 6 times giving
\be\label{Tilbsw}
T_{\rm i}\gtrsim [4\,(2.5)\times 10^9 - 2\,(1.5)\times 10^{10}]\,{\rm GeV} \, .
\ee
This is another interesting result showing how the support of
neutrino mixing data to the range $\mt\sim {\cal O}(m_{\rm sol},m_{\rm atm})$
not only makes leptogenesis working in a simple
and predictive way but also how the loss in the efficiency is compensated
by a non relativistic production of the final asymmetry such that the lower
bound on the reheating temperature gets just slightly more restrictive compared
to the small $\mt$ range. Note that in case of modifications of the theoretical
assumptions such that the maximum baryon asymmetry
$\eta_{B}^{\rm max}\rightarrow \xi\,\eta_B^{\rm max}$, one has correspondingly
$M_1^{\rm min},T_i^{\rm min}\rightarrow M_1^{\rm min}/\xi,T_i^{\rm min}/\xi$.

\subsection{Upper bound on the absolute neutrino mass scale}

For large values of the absolute neutrino mass scale the $\Delta W$
wash-out term cannot be neglected. The final efficiency factor can be
calculated in the approximation that $\Delta W$ starts to be effective
for $z>z_B$, when the asymmetry generation from decays already stopped.
This is a very good approximation in the strong wash-out regime and
since $\mt\geq m_1$ it does not introduce any restriction for
$m_1\gtrsim m_{\star} \simeq 10^{-3}\,{\rm eV}$. Within this approximation
one has simply
\be
\k_{\rm f}(\mt,\,M_1\mb^2)=\kappa_{\rm f}(\mt)
\,e^{- {\o\over z_B}\,
\left({M_1\over 10^{10}\,{\rm GeV}}\right)\,
\left({\mb\over {\rm eV}}\right)^2}\, ,
\ee
where $\k_{f}(\mt)$ is the efficiency factor calculated in the regime of small
neutrino masses neglecting the $\Delta W$ term. We use the simple limit
$k_{\rm f}=2/(z_B\,K)$ that corresponds to neglect, conservatively, the contribution of scatterings
to the wash-out and approximately to the upper values in the Eq. (\ref{plaw}).
A search of the peak values $(M_1,\mt)$ for which the maximum baryon
asymmetry $\eta_B^{\rm max}$ has an absolute maximum yields
\be
{\eta_{B}^{\rm peak}(m_1)\over \eta_B^{CMB}} \simeq {2\over 3^{7/2}}\,
{\chi\,m_*\,\xi \over m_i^4} \;,
\ee
with the constant $\chi \simeq 1.6\,{\rm eV}^3$. Hence
the CMB bound implies an interesting constraint on the neutrino mass,
given by $m_i< 0.1\,{\rm eV}$ \cite{bdp2,bdp3}.
A precise calculation has to take into account the running of neutrino masses \cite{ratz}.
The atmospheric neutrino mass scale
at temperatures $T \sim 10^{13}\,{\rm GeV}$ is higher than at zero temperature.
On the other hand the bound on neutrino masses that is obtained at large temperatures
gets lower when calculated at low temperatures. This second effect is dominant and thus
the account of neutrino mass running will make the neutrino mass bound more restrictive.
 The smallest effect is obtained for an Higgs mass $M_h \simeq 150\,{\rm GeV}$
and makes the bound $20\%$ more stringent. Taking into account this effect one then
obtains the $3\sigma$ bound \cite{pedestrians}
\be\label{mbound}
m_i < 0.12\,{\rm eV}\,\xi^{1/4}  \, .
\ee
Note however that  a (1 figure) bound $m_i < 0.1\,{\rm eV}$ conservatively
accounts for the theoretical uncertainties
\footnote{If instead of the upper values we were using the central (lower) values
in the Eq. (\ref{plaw}), then the bound would have been $\sim 0.005\,(0.01)\,{\rm eV}$ more
stringent. As we said already, these values could arise from a possible contribution to the
wash-out from scatterings or for the account of spectator processes. Note also that
we used the latest published results on $\Delta m^2_{\rm atm}$ \cite{atm}.
Preliminary results from the SuperKamiokande collaboration \cite{talk} find
$\Delta m^2_{\rm atm}=(1.3-3.0)\times 10^{-3}\,{\rm eV}^2$ at $90\%$ c.l.
(best fit $\Delta m^2_{\rm atm}=2.0\times 10^{-3}\,{\rm eV}^2$),
from which, at 1$\sigma$, $m_{\rm atm}=(0.045\pm 0.006)\,{\rm eV}$, implying a
$\sim 0.01\,{\rm eV}$ more stringent bound $m_i< 0.11\,{\rm eV}$.
In \cite{gnrrs} a bound $m_i< 0.15\,{\rm eV}$, when this second value of
$m_{\rm atm}$ is employed, has been obtained. A difference $\sim 0.02\,{\rm} eV$
can be ascribed to a different estimation of the running of neutrino masses.
The remaining $0.02\,{\rm eV}$ difference can be safely included in
the account of the theoretical uncertainty.}.
As defined in the previous subsection, a value of $\xi\neq 1$ describes
a possible variation of the maximum baryon asymmetry
in the case of modified theoretical assumptions, like for example in the supersymmetric case
that will be studied in the next subsection, or in the presence of possible different
effects, like an enhancement of the $C\!P$ asymmetry due to a degenerate heavy neutrino
spectrum \cite{CPas} that would relax the bounds \cite{bdp3,resonant,hambye},
or simply for a variation of the input values
of the experimental quantities (note that $\xi\propto m_{\rm atm}^2/\eta_B^{CMB}$).
However the strong suppression of
the baryon asymmetry for an increasing absolute neutrino mass scale
($\eta_B\propto 1/m_i^4$) makes the bound quite stable \cite{bdp3}.
It is important to realize that the bound can be evaded but not trivially.
This means that a measurement of a value of the absolute neutrino mass scale above the
leptogenesis bound will necessarily imply some drastic modifications of the minimal
leptogenesis scenario. These include particular neutrino models
within the simple seesaw formula \cite{bdp3,hambye},
non thermal leptogenesis scenarios \cite{nont} or a non minimal seesaw formula,
like that one arising in theories with a triplet Higgs \cite{hs}.

\subsection{The supersymmetric case}

Leptogenesis can be also studied within the
minimal supersymmetric standard model (MSSM) \cite{sacha,plum2,gnrrs}.
In this case the asymmetry is generated not only from the
$N_1$ decays  but also from the decays of their scalar partners,
the $\widetilde{N_1^c}$'s and their antiparticles $\widetilde{N_1^c}^{\dagger}$,
with the same mass $M_1$. Since the decay width
and thus also the inverse decay wash-out term are the same, these yield
an additional equal contribution \footnote{The discussion in this subsection is from \cite{moriond}.}.
Therefore, from a thermodynamical point of view, the $N_1$'s and the
$\widetilde{N}_1$'s will play the same role and it is simply like if the
`X-abundance' gets doubled. We will still track the $B-L$ asymmetry in the co-moving volume
containing, on average in ultra-relativistic thermal equilibrium,
one RH neutrino $N_1$ and thus now also one $\widetilde{N_1}$.
In this way we will still write the final baryon asymmetry as
in the Eq. (\ref{etaB}) but now
$d=2\,{a_{\rm sph}/N_{\gamma}^{\rm rec}}$,
with the additional factor 2 taking into account the contribution from the
$\widetilde{N_1^c}$'s
\footnote{The other possibility would have been to choose a halved
co-moving volume in a way that the factor $2$ was absorbed in $N_{\gamma}^{\rm rec}$.}.
 Let us analyze how the different quantities involved in the calculation of the
 final asymmetry get modified from the SM to the MSSM case. To this aim
it will prove convenient to introduce the variations
\be
\xi_X\equiv {X^{MSSM}\over X^{SM}} \,
\ee
for any quantity $X$.
The sphaleron conversion coefficient is given by $a_{\rm sph}=8/23$ and thus
it is almost unchanged. The number of degrees of freedom is given by
$g_{MSSM}=915/4$ and this approximately doubles the number of photons
at recombination given by
\be
(N_{\gamma}^{\rm rec})^{MSSM}=4\,{g_{MSSM}\over 3\,g_{\rm rec}}
={3355\over 43}\simeq 78 \, .
\ee
In this way one gets $d^{MSSM}\simeq 0.89\times 10^{-2}$, almost unchanged compared to the SM case
($\xi_d\simeq 0.92$).
The $C\!P$ asymmetry in the MSSM case is given by \cite{CPas}
\be\label{eps1gMSSM}
\varepsilon_1 \simeq  -{1\over 8\pi}
\sum_{i=2,3}\,{{\rm Im}\,\left[(h\,h^{\dagger})^2_{i1}\right]\over (h\,h^{\dagger})_{11}} \,\times\,
\left[f_V\left({M^2_i\over M^2_1}\right)+f_S\left({M^2_i\over M^2_1}\right)\right] \, ,
\ee
where
\be
f_V(x)=\sqrt{x}\,\ln\left({1+x\over x}\right)
\ee
and
\be
f_S(x)={\sqrt{x}\over 1-x}  \, .
\ee
In the limit $x\gg 1$, corresponding to have a mild RH neutrinos mass hierarchy
with $M^2_i \gg M^2_1$ and barring strong phase cancellations, one has
\be
f_V(x)+f_S(x)\longrightarrow {3\over \sqrt{x}}
\ee
and thus, compared to the SM case (cf. (\ref{eps1g}) and (\ref{flimit})),
the absolute value of the $C\!P$ asymmetry and hence also of its maximum
$\ve_1^{\rm max}$  get doubled ($\xi_{\varepsilon}=2$).

Let us now calculate how the efficiency factor at small values of
$M_1\,\bar{m}^2$ gets modified in the MSSM case.
 We have seen that SM leptogenesis
 is well approximated, in the strong wash-out regime,
 by the simple decays plus inverse decays picture.
 It is then interesting to study how the results change within such a simple picture.
Since the $N_1$'s can now decay in two new channels
($N_1\rightarrow\tilde{l}\bar{\tilde{h}},\tilde{l}^{\dagger}\tilde{h}$)
that give exactly the same contribution to the decay width as the other two standard ones,
this gets doubled compared to the standard case (cf. (\ref{Gamma})).
This makes lifetime shorter and the inverse decays wash out rate
stronger. However the increase of the degrees of freedom makes the expansion
faster and this partially compensates.
Recalling the definition of the {\em equilibrium neutrino mass}, Eq. (\ref{mstar}),
one has simply
\be
\xi_{m_{\star}}=
{\sqrt{\xi_{g_{\star}}}\over \xi_{\Gamma_D^{\rm rest}}}=
{1\over 2}\,\sqrt{915\over 427} \simeq  0.73   \, ,
\ee
implying $m_{\star}^{\rm MSSM}\simeq 0.8\times\,10^{-3}\,{\rm eV}$.
Therefore now one has
that the transition to the strong wash-out regime occurs
for slightly smaller values of $\mt$ (about $1/\sqrt{2}$ smaller)
and thus for the decay parameter one has $\xi_K\simeq\sqrt{2}$.
The range of $K$ values favored by neutrino mixing
data will thus be given by
\be\label{KleprangeMSSM}
{\cal O}(K_{\rm sol}^{\rm MSSM}\simeq 10)\lesssim K^{MSSM}
\lesssim {\cal O}(K_{\rm atm}^{MSSM}\simeq 65) \, ,
\ee
to be compared with the Eq. (\ref{Kleprange}) in the SM case.
In the strong wash out regime the efficiency factor, calculated in the
decays plus inverse decays picture, will be still given by
the expression Eq. (\ref{kfth}).
For $K$ in the range (\ref{KleprangeMSSM}) one has $\k_{\rm f}\propto K^{-1.15}$
(cf. (\ref{kfpllep})) and thus
\be\label{xikf}
\xi_{k_{\rm f}} \simeq \xi_{m_{\star}}^{1.15} \simeq {1\over\sqrt{2}}  \, .
\ee
Assuming that the effect of scatterings in the MSSM goes in the
same direction as in the SM and thus that the result
$\xi_{k_{\rm f}}\simeq 1/\sqrt{2}$ holds approximately also when
scatterings are included, one can write the analogous of the
Eq. (\ref{kf}) in the MSSM case
\be\label{plawMSSM}
\k_{\rm f}^{MSSM}=(1.5\pm 0.7)\,10^{-2}\,\left({{\rm 10^{-2}\,eV}\over\mt}\right)^{1.1\pm 0.1}  \, .
\ee
A more detailed analysis is needed to verify the role of scatterings in the MSSM.
Note that again the upper values of this range correspond to
the power law (\ref{kfpllep}) translated in the MSSM case
(i.e. by replacing $m_{\star}^{SM}\rightarrow m_{\star}^{MSSM}$).

Let us now investigate the consequences for the lower bound on $M_1$.
The Eq. (\ref{lbM1}) will now become
\bea\label{lbM1MSSM}
M_1 \geq M_1^{\rm min} &=& {1\over d}\,
{8\,\pi\over 3}\,{v^2\over m_{\rm atm}}\,
{\eta_B^{CMB}\over\k_{\rm f}} \NO\\
&\simeq& 3.5\times 10^{8}\,{\rm GeV} \left(\eta_B^{CMB}\over
6\times 10^{-10}\right)
\left(0.05\,{\rm eV}\over m_{\rm atm}\right)\,\k_{\rm f}^{-1} \;.
\eea
From this expression one obtains, for the case of initial thermal abundance, zero initial abundance
and in the strong wash-out regime respectively, the constraints
\be\label{M1lbzaMSSM}
M_1 \geq
(3.7\pm 0.4)\times 10^8\,{\rm GeV}
\gtrsim 2.5\times 10^8\,{\rm GeV} \, ,
\ee
\be\label{M1lbtaMSSM}
M_1 \geq (1.7 \pm 0.2) \times 10^9\,{\rm GeV}
\gtrsim 1.1 \times 10^9\,{\rm GeV} \,
\ee
and
\be\label{M1lbswMSSM}
M_1\gtrsim [2.4\,(1.6)\pm 0.2\,(0.15)] \times 10^{10}\,{\rm GeV}\,
\left(\widetilde{m}_1\over 10^{-2}\,{\rm eV} \right)^{1.1(1.2)}
\gtrsim [1.5\,(0.9) - 10\,(8)]\times  10^{10}\,{\rm GeV} \, ,
\ee
where we have used the central (upper) value in the Eq. (\ref{plawMSSM}).
These have to be compared with the constraints
obtained in the SM case (cf. Eq.'s (\ref{M1lbza}),
(\ref{M1lbta}) and (\ref{M1lbsw})).
In the first two cases the constraints are approximately twice
looser, because of the $C\!P$ asymmetry enhancement,
while in the case of strong wash out
one has $\xi_{M_1^{\rm min}}\simeq
(\xi_{\varepsilon}\,\xi_{\k_f})^{-1}\simeq 0.7$ for the central value,
while the $3\sigma$ bound remains practically unchanged because the
experimental error gets reduced.
For the lower bound on the initial temperature, corresponding to the reheating temperature
within inflation, the same considerations as in the SM case hold. For
$\mt\lesssim 10^{-3}\,{\rm eV}$ the same lower bounds valid for $M_1$
apply approximately also to $T_{\rm i}$.

In the relevant strong wash-out
regime for $K\gtrsim 4$, corresponding to $\mt\gtrsim 3.2\times 10^{-3}\,{\rm eV}$,
the relaxation compared to the $M_1$ lower bound is larger than twice. In this way
one obtains conservatively, using the upper values for $\k_{\rm f}$ (cf. (\ref{plawMSSM})),
\be\label{Timin}
T_{i}\gtrsim 1.5\times \,10^{9}\,{\rm GeV} \, .
\ee
This is an appropriate conservative value for a generical comparison with the upper bounds
on the reheating temperature that arise by imposing that a gravitino thermal
production is not in conflict with cosmological observations
(see \cite{focus} for a recent discussion and references).
For a more  precise comparison one should calculate the lower bound on $T_{\rm i}$
for a specific value of $\mt$.
For example, in the range $m_{\rm sol}\lesssim \mt \lesssim m_{\rm atm}$, the relaxation
compared to the $M_1$ lower bound is practically the same as in the SM case (between 4 and 6 times)
and thus one gets
\be
T_i \gtrsim [3.5\,(2)\times 10^9 - 2\,(1.5)\times  10^{10}]\,{\rm GeV}  \, .
\ee
Note that these analytic results are in good agreement with the
numerical ones in \cite{gnrrs}.

Let us now study how the {\em upper bound on the neutrino masses}
gets modified. This can be easily done calculating the value of
the total variation of the final asymmetry given by \cite{pedestrians}
\be
\xi={\xi_{\varepsilon}\,\xi_{\k_{\rm f}}\over \xi_{\o}} \, ,
\ee
where $\xi_{\o}$ is the variation of the wash-out term from
$\Delta L=2$ processes that is crucial for the determination of the
neutrino mass bound and that can be expressed through the parameter
$\o$ in the Eq. (\ref{nonr}). We have already seen that
$\xi_{\varepsilon}= 2$. Moreover since the peak of the asymmetry lies
in the strong wash-out regime we can also use the result Eq. (\ref{xikf})
for $\xi_{\k_{\rm f}}$. Therefore, we miss only to determine $\xi_{\o}$.
There are two effects to be considered.
The first is the increase of the number of degrees of freedom that
speeds up the expansion reducing the efficiency of wash-out processes.
The second is the presence of new different additional
$\Delta L=2$ processes and this clearly strengthens the rate $\Gamma_{\Delta L=2}$.
From the expressions given in \cite{plum2}
one can find $\xi_{\Gamma_{\Delta L=2}}=5/3$
and thus $\xi_{\o}=\xi_{\Gamma_{\Delta L=2}}/\sqrt{\xi_{g_{\star}}}\simeq 5/(3\,\sqrt{2})$.
 Putting all together one finds $\xi \simeq 6/5$ and, when the running of neutrino masses is neglected,
the bound is about $5\%$ more relaxed compared to the SM case, namely
$m_i^{MSSM}\lesssim 0.16\,{\rm eV}$. The effect of running of neutrino masses, as in the SM,
goes into the direction to make the bound more stringent. However the
effect can be as small as $\sim 7\%$ (for ${\rm tan\b \sim 10}$ \cite{ratz}),
roughly half than in the SM case, and
thus one obtains in the end (at $3\,\sigma$)
\be\label{mimaxMSSM}
m_i^{MSSM} < 0.15\,{\rm eV} \, .
\ee

\subsection{A `too-short-blanket problem'}

If one requires that $M_1$ is lower than a certain cut-off value $M_1^{\star}$ then
the upper bound on the neutrino masses becomes more stringent \cite{bdp3}.
Such a cut-off can either arise directly from neutrino models \cite{smirnov}
 or indirectly from
an upper bound on the reheating temperature $T_R^{\star}$. In this second case one has
$M_1^{\star}\simeq z_{\star}\,T_R^{\star}$ where $z_{\star}\simeq 1$ in the weak wash-out regime
and $z_{\star}\simeq z_B-2$ in the strong wash-out regime
\footnote{The value of $z_{\star}$ has to be evaluated
for that particular value of $\mt$ that maximizes the asymmetry.}.
It is then quite interesting to study the
dependence $m_1^{\rm bound}(M_1^{\star})$, or equivalently $m_1^{\rm bound}(T_R^{\star})$.
This is done in detail in \cite{prep}, here we just sketch some general features and results.
For definiteness we will refer to the supersymmetric case, since in this case the
avoidance of the gravitino problem implies an upper bound on the reheating temperature.

First, note that if $T_{R}^{\star}<T_R^{\rm min}\simeq 10^9\,{\rm GeV}$ (cf.(\ref{Timin})),
then simply there is no allowed value for $m_1$. Moreover until one has $m_1^{\rm bound}\ll m_{\rm atm}$,
there is a strong dependence on $T_R^{\star}$, since the maximum baryon asymmetry
grows linearly with $T_R^{\star}$ while is very slightly dependent on $m_1$.
This means  that the function $m_1^{\rm bound}(T_R^{\star})$
has a vertical asymptote in $T_R^{\rm min}\simeq 10^9\,{\rm GeV}$.
For $m_1^{\rm bound}\sim m_{\rm atm}$ the suppression factor $\beta_{\rm max}(m_1)=m_{\rm atm}/(m_1+m_3)$
in the $C\!P$ asymmetry and the loss in the efficiency for $\mt>m_1\gtrsim 10^{-3}\,{\rm eV}$
compensate the increase of $T_R^{\star}$ and the growth of
$m_1^{\rm bound}$, for increasing $T_R^{\star}$, slows down and eventually,
for $T_R^{\star}\geq T_R^{\rm peak}\simeq 3\times 10^{12}\,{\rm GeV}$,
saturates to its maximum value (cf.(\ref{mimaxMSSM})) and stays constant
\footnote{Note that at the peak one has $z_B\simeq 10$ and $M_1^{\rm peak}\simeq 2\times 10^{13}\,{\rm GeV}$
and thus $T_R^{\rm peak}\simeq M_1^{\rm peak}/8 \simeq 3\times 10^{12}\,{\rm GeV}$.}.
Values  $T_R^{\star}\sim 10^{10}\,{\rm GeV}$ are particularly interesting since they
correspond to maximum allowed values from gravitino problem arguments.
In this case one expects $m_1\sim m_{\rm atm}\gg 10^{-3}\,{\rm eV}$ and,
since $\mt\geq m_1$, one can use the strong wash-out limit for the
final efficiency factor $\k_{\rm f}\simeq 2/(K\,z_B)$.
Moreover
the wash-out factor $\Delta W$ from $\Delta L=2$ processes can be neglected.
In this case it is possible to show the following approximate bound
\be
{m_1\over m_{\rm atm}} \lesssim {A\over\sqrt{1+2\,A}} \,\;\; \;\; \mbox{with} \;\;\;\; A \simeq
{0.2\over\xi_{\eta}\,\xi_{\rm atm}}\,{T_R^{\star}\over 10^{10}\,{\rm GeV}} \, ,
\ee
where $\xi_{\eta}=\eta_B^{CMB}/6\times 10^{-10}$ and $\xi_{\rm atm}=m_{\rm atm}/0.051\,{\rm eV}$.

Let us consider two examples using $\xi_{\eta}=\xi_{\rm atm}=1$.
For an upper bound $T_R^{\star}=3\times 10^{10}\,{\rm GeV}$ one finds
$m_1\lesssim 0.4\,m_{\rm atm}\simeq 0.02\,{\rm eV}$, implying $m_3\lesssim 0.055\,{\rm eV}$.
For $T_R^{\star}=10^{11}\,{\rm GeV}$ one finds
$m_1\lesssim 0.9\,m_{\rm atm}\simeq 0.045\,{\rm eV}$, implying $m_3\lesssim 0.07\,{\rm eV}$.

This exercise shows that it is difficult to conciliate
reheating temperatures close to the minimum allowed one ($T_R\lesssim 10^{10}\,{\rm GeV}$)
and at the same time to make thermal leptogenesis compatible with quasi-degenerate neutrino
masses by evading the upper bound: the two things go into opposite
directions, a typical {\em too-short-blanket problem}.
There are two interesting consequences. The first is that
if a stringent upper bound on the reheating temperature
is placed, like $T_R\lesssim 3\times 10^{10}\,{\rm GeV}$, then it becomes difficult to
evade the bound invoking a quasi-degenerate {\em heavy} neutrino spectrum since the bound still falls in
a transition region where light neutrinos exhibit a partial hierarchy.
Vice versa if one requires quasi-degenerate light neutrinos to be compatible with the minimal thermal
leptogenesis scenario then the problem of a large minimum reheating temperature gets exacerbated. Indeed,
it is difficult in this case to avoid high values $T_R\gtrsim 10^{11}\,{\rm GeV}$,
unless one invokes a strong degenerate heavy neutrino spectrum such to have a resonant enhancement of
the asymmetry \cite{resonant,hambye}. Anyway further investigations are needed
to understand the exact conditions on the degeneracy of the heavy neutrino spectrum and how they
depend on a cut-off on $T_R$ or directly on $M_1$.







\section{Final discussion}

Leptogenesis is a specific realization of the simplest and oldest
baryogenesis class of models where the asymmetry is generated
from heavy particle decays. Its minimal version, thermal leptogenesis,
is based crucially on neutrino properties and because of the
great experimental  neutrino physics achievements it became in the
last year a testable model. The decay parameter, the key quantity in
models of baryogenesis from heavy particle decays, is a quantity closely
related to neutrino masses. This cannot be exactly determined from data
but there is an emerging favored range of values,
$K_{\rm lep}\sim 5-50$, that implies just a small departure from thermal
equilibrium, however large enough to explain the observed value of the asymmetry
and moreover with some nice consequences. The predicted baryon asymmetry is independent on the
initial conditions, both on the initial value of the asymmetry and on the initial
number of decaying RH neutrinos. Moreover, the theoretical uncertainties
are minimized and the final asymmetry is predicted with a precision that is
within half order of magnitude. These features can be synthesized
saying that thermal leptogenesis predictions are quite stable and model independent,
a picture that resembles very closely the Standard Big Bang Nucleosynthesis in
predicting the primordial nuclear abundances.
The drawback is that values of $K\sim 10$ determine a loss in the efficiency between one and two orders
of magnitude. This has to be compensated by an increase of the $C\!P$ asymmetry of the same amount,
implying a more stringent lower bound on $M_1$. On the other hand we have seen how
in the strong wash out regime the temperature of baryogenesis gets much smaller than
$M_1$ and this relaxes the lower bound on the reheating temperature compared
to the lower bound on $M_1$ of a factor $\sim 5$. Therefore, there seems to be an
intriguing conspiracy between neutrino mixing data and the explanation of
the observed baryon asymmetry.
Actually considerations on the maximum allowed value of the effective neutrino mass
show that the conspiracy is even deeper \cite{focus} and
future experimental information on the absolute neutrino mass scale could
give a further support.

If the leptogenesis upper bound
on the absolute neutrino mass scale, $m_i < 0.1\,{\rm eV}$,
will be fully tested with cosmology, neutrinoless double beta decay
and Tritium beta decay experiments, then  thermal leptogenesis can work in its minimal way.
On the other hand, if neutrino masses higher than $0.1\,{\rm eV}$ will be found, then
this can be either regarded as the effect of the existence of some level of degeneracy in the
heavy neutrino spectrum, to be understood whether easily realized or not within the simple
seesaw mechanism, or, more likely, as a drastic departure from the minimal thermal
leptogenesis picture, at the expense of predictivity. In any case it should be clear,
from the discussion on the `too-short-blanket problem', that the two statements for which
thermal leptogenesis requires dangerously large reheating temperatures within the
supersymmetric framework and that the neutrino mass bound can be evaded within minimal thermal
leptogenesis, can be very difficultly made compatible with each other.

Another interesting aspect is that if the lightest neutrino mass $m_1$ will be found to be higher than
$m_{\star}\simeq 10^{-3}\,{\rm eV}$, then it will be possible to conclude model independently
that thermal leptogenesis lies in the strong wash-out regime and in this way all
pieces of the experimental information will have fitted within the theoretical best expectations.
Therefore, if the absolute neutrino mass scale will be found to lie within the
window $(10^{-3}- 10^{-1})\,{\rm eV}$, the picture will receive further strong support
from the data.

\vspace{3mm}

\noindent {\bf Acknowledgments}

\noindent
It is a pleasure to thank  W. Buchm\"{u}ller and M. Pl\"{u}macher for a fruitful collaboration
and valuable comments, J. Pati and T. Hambye for interesting discussions and the organizers of
NO-VE03 and Moriond Electroweak 2004 for such nice and interesting meetings.


\begin{thebibliography}{99}

\bibitem{nove}
P.~Di Bari,
{\em Leptogenesis and neutrino mixing data},
Proceedings of the {\em Second International Workshop
on Neutrino Oscillation in Venice}, ed. Baldo Ceolin (3-5 December 2003, Venice), p.497.

\bibitem{moriond}
P.~Di Bari,
{\em Leptogenesis neutrino mass bound}, to appear in the
Proceedings of the {\em 39th Rencontres de Moriond on
Electroweak Interactions and Unified Theories},
(March 2004, La Thuille).

\bibitem{pedestrians}
W.~Buchmuller, P.~Di Bari and M.~Plumacher,
to appear in Annals of Physics, arXiv:hep-ph/0401240.

\bibitem{Glashow}
A.~G.~Cohen, A.~De Rujula and S.~L.~Glashow,
Astrophys.\ J.\  {\bf 495} (1998) 539 [arXiv:astro-ph/9707087].

\bibitem{WMAP}
WMAP Collaboration, D.~N.~Spergel {\it et al.},
Astrophys.\ J.\ Suppl.\  {\bf 148} (2003) 175.

\bibitem{SDSS}
Max Tegmark {\it et al.}, astro-ph/0310723.

\bibitem{Cyburt}
R.~H.~Cyburt, B.~D.~Fields and K.~A.~Olive,
New Astron.\  {\bf 6} (2001) 215 [arXiv:astro-ph/0102179];
D.~Kirkman, D.~Tytler, N.~Suzuki, J.~M.~O'Meara and D.~Lubin,
arXiv:astro-ph/0302006.

\bibitem{Sakharov}
A.~D.~Sakharov,
Pisma Zh.\ Eksp.\ Teor.\ Fiz.\  {\bf 5} (1967) 32.

\bibitem{reviews}
A.~D.~Dolgov,
Phys.\ Rept.\  {\bf 222} (1992) 309;
M.~Dine and A.~Kusenko,
Rev.\ Mod.\ Phys.\  {\bf 76} (2004) 1 [arXiv:hep-ph/0303065];
A.~Riotto and M.~Trodden,
Ann.\ Rev.\ Nucl.\ Part.\ Sci.\  {\bf 49} (1999) 35 [arXiv:hep-ph/9901362].

\bibitem{fy}
M.~Fukugita and T.~Yanagida,
Phys.\ Lett.\ B {\bf 174} (1986) 45.

\bibitem{kt83}
For a review see E.~W.~Kolb and M.~S.~Turner,
Ann.\ Rev.\ Nucl.\ Part.\ Sci.\  {\bf 33} (1983) 645.

\bibitem{kt}
E.~W.~Kolb, M.~S.~Turner, {\it The Early Universe}, Addison-Wesley,
New York, 1990.

\bibitem{sphalerons}
G.~t'Hooft, \prl{37}{1976}{8};
V.~A.~Kuzmin, V.~A.~Rubakov,  M.~E.~Shaposhnikov, \pl{155}{1985}{36}.

\bibitem{luty}
M.~A.~Luty, \pr{45}{1992}{455}.

\bibitem{plum}
M.~Pl\"umacher, Z.~Phys.~{\bf C\ 74} (1997) 549.

\bibitem{bcst}
R.~Barbieri, P.~Creminelli, A.~Strumia and N.~Tetradis,
Nucl.\ Phys.\ B {\bf 575} (2000) 61
[arXiv:hep-ph/9911315].

\bibitem{bdp1}
W.~Buchm\"uller, P.~Di~Bari, M.~Pl\"umacher, \np{643}{2002}{367}.

\bibitem{dolgov}
A.D.~Dolgov, Sov. J. Nucl. Phys. {\bf 32}, 831 (1980);
E.~W.~Kolb, S.~Wolfram, \np{172}{1980}{224}.

\bibitem{bdp3}
W.~Buchm\"uller, P.~Di~Bari, M.~Pl\"umacher, \np{665}{2003}{445}.

\bibitem{Fry}
J.~N.~Fry, M.~S.~Turner, \pr{24}{1981}{3341}.

\bibitem{yasu}
H.~B.~Nielsen and Y.~Takanishi,
Phys.\ Lett.\ B {\bf 507} (2001) 241 [arXiv:hep-ph/0101307].


\bibitem{casas}
J.~A.~Casas and A.~Ibarra,
Nucl.\ Phys.\ B {\bf 618} (2001) 171 [arXiv:hep-ph/0103065].



\bibitem{hama}
M.~Fujii, K.~Hamaguchi, T.~Yanagida, \pr{65}{2002}{115012}.


\bibitem{atm}
M.~H.~Ahn et al., K2K Collaboration,
\prl{90}{2003}{041801};\\
M. Shiozawa et al., SK Collaboration in {\em Neutrino 2002}, Proc. to appear;
G.~L.~Fogli, E.~Lisi, A.~Marrone, D.~Montanino, A.~Palazzo, A.~M.~Rotunno,
hep-ph/0310012.

\bibitem{solar}
Q.R. Ahmad {\it et al}, SNO Collaboration, nucl-ex/0309004;\\
K.~Eguchi {\it et al.}, KamLAND Collaboration,
\prl{90}{2003}{021802}

\bibitem{ir}
See for example:
S.~Pascoli, S.~T.~Petcov and W.~Rodejohann,
Phys.\ Rev.\ D {\bf 68} (2003) 093007
[arXiv:hep-ph/0302054];
A.~Ibarra and G.~G.~Ross,
Phys.\ Lett.\ B {\bf 575} (2003) 279 [arXiv:hep-ph/0307051].


\bibitem{gnrrs}
G.~F.~Giudice, A.~Notari, M.~Raidal, A.~Riotto and A.~Strumia,
arXiv:hep-ph/0310123.

\bibitem{pilun}
A.~Pilaftsis and T.~E.~J.~Underwood,
arXiv:hep-ph/0309342.

\bibitem{spectator}
W.~Buchmuller and M.~Plumacher,
Phys.\ Lett.\ B {\bf 511} (2001) 74
[arXiv:hep-ph/0104189].

\bibitem{fy2}
M.~Fukugita and T.~Yanagida,
Phys.\ Rev.\ D {\bf 42} (1990) 1285.

\bibitem{CPas}
M.~Flanz, E.~A.~Paschos, U.~Sarkar, \pl{345}{1995}{248};
\pl{384}{1996}{487} (E);
L.~Covi, E.~Roulet, F.~Vissani, \pl{384}{1996}{169};
W.~Buchm\"uller, M.~Pl\"umacher, \pl{431}{1998}{354}.


\bibitem{hambye}
T.~Hambye, Y.~Lin, A.~Notari, M.~Papucci and A.~Strumia,
arXiv:hep-ph/0312203.

\bibitem{gkr}
G.~F.~Giudice, E.~W.~Kolb and A.~Riotto,
Phys.\ Rev.\ D {\bf 64} (2001) 023508
[arXiv:hep-ph/0005123].



\bibitem{fred}
W.~Buchmuller and S.~Fredenhagen,
Phys.\ Lett.\ B {\bf 483} (2000) 217 [arXiv:hep-ph/0004145].

\bibitem{di}
S.~Davidson, A.~Ibarra, \pl{535}{2002}{25}.

\bibitem{prep}
P.~Di Bari, in preparation.

\bibitem{asaka}
T.~Asaka, K.~Hamaguchi, M.~Kawasaki and T.~Yanagida,
Phys.\ Lett.\ B {\bf 464} (1999) 12
[arXiv:hep-ph/9906366];
K.~Hamaguchi, H.~Murayama, T.~Yanagida, \pr{65}{2002}{043512}.


\bibitem{bdp2}
W.~Buchmuller, P.~Di Bari and M.~Plumacher,
Phys.\ Lett.\ B {\bf 547} (2002) 128
[arXiv:hep-ph/0209301].

\bibitem{bdp3}
W.~Buchm\"uller, P.~Di~Bari, M.~Pl\"umacher, \np{665}{2003}{445}.

\bibitem{ratz}
S.~Antusch, J.~Kersten, M.~Lindner and M.~Ratz,
Nucl.\ Phys.\ B {\bf 674} (2003) 401 [arXiv:hep-ph/0305273].

\bibitem{talk}
K. Okumura for the SuperKamiokande collaboration,
Proceedings of the {\em Second International Workshop
on Neutrino Oscillation in Venice}, ed. Baldo Ceolin (3-5 December 2003, Venice), p.129.


\bibitem{resonant}
A.~Pilaftsis and T.~E.~J.~Underwood,
arXiv:hep-ph/0309342.

\bibitem{nont}
G.~Lazarides, Q.~Shafi, \pl{258}{1991}{305};
H.~Murayama, T.~Yanagida, \pl{322}{1994}{349};
G.~F.~Giudice, M.~Peloso, A.~Riotto and I.~Tkachev, JHEP {\bf 9908} (1999) 014.


\bibitem{hs}
See T.~Hambye and G.~Senjanovic,
Phys.\ Lett.\ B {\bf 582} (2004) 73
[arXiv:hep-ph/0307237];
S.~Antusch and S.~F.~King,
arXiv:hep-ph/0405093
and references therein.


\bibitem{sacha}
B.~A.~Campbell, S.~Davidson and K.~A.~Olive,
Nucl.\ Phys.\ B {\bf 399} (1993) 111 [arXiv:hep-ph/9302223].

\bibitem{plum2}
M.~Plumacher,
Nucl.\ Phys.\ B {\bf 530} (1998) 207
[arXiv:hep-ph/9704231].

\bibitem{focus}
W.~Buchmuller, P.~Di Bari and M.~Plumacher, hep-ph/0406014.


\bibitem{smirnov}
See for example:
G.~C.~Branco, R.~Gonzalez Felipe, F.~R.~Joaquim and M.~N.~Rebelo,
Nucl.\ Phys.\ B {\bf 640} (2002) 202;
E.~K.~Akhmedov, M.~Frigerio and A.~Y.~Smirnov, JHEP {\bf 0309} (2003) 021.



\end{thebibliography}
\end{document}